\documentclass{aastex631}

\usepackage{graphicx,subfigure,color,amsmath,pifont,multirow,threeparttable}
\usepackage{acronym}
\usepackage{mathrsfs}
\usepackage{gensymb}
\usepackage{hyperref}

\shorttitle{Post-merger Magnetar-Driven Poynting Jet and Pulsar Wind Nebula}
\shortauthors{Li et al.}
\begin{document}
\title{Multi-wavelength Emission for a Post-merger Magnetar: The Magnetar-Driven Poynting Jet and Its Associated Pulsar Wind Nebula}
\correspondingauthor{Da-Bin Lin}
\email{lindabin@gxu.edu.cn}
\author{Yun-Peng Li}
\affiliation{Guangxi Key Laboratory for Relativistic Astrophysics, School of Physical Science and Technology, Guangxi University, \\Nanning 530004, China}
\author{Da-Bin Lin}
\affiliation{Guangxi Key Laboratory for Relativistic Astrophysics, School of Physical Science and Technology, Guangxi University, \\Nanning 530004, China}
\author{Ning-Yuan Zhang}
\affiliation{Guangxi Key Laboratory for Relativistic Astrophysics, School of Physical Science and Technology, Guangxi University, \\Nanning 530004, China}
\author{En-Wei Liang}
\affiliation{Guangxi Key Laboratory for Relativistic Astrophysics, School of Physical Science and Technology, Guangxi University, \\Nanning 530004, China}

\begin{abstract}
A newborn, rapidly rotating magnetar may form in a binary neutron star merger and drive a Poynting-flux-dominated relativistic jet. As the jet propagates outward, a forward shock (FS) and a reverse shock (RS) are formed, inflating a pulsar wind nebula (PWN) between them. We present a systematic study of the emission from both the PWN and the jet, whose magnetic energy is subject to dissipation. By following the dynamics of the jet–ejecta–PWN system, we find that, in most cases, the RS is long-lived: it first lags behind the contact discontinuity and eventually coincides with both the contact discontinuity and the FS after the jet breakout into the external medium. As a result, the emission exhibits a characteristic temporal evolution. Depending on the optical depth, the emission is initially dominated by thermal radiation from the optically thick ejecta, then by a jet-powered X-ray plateau once the system becomes optically thin, and finally by synchrotron and inverse-Compton radiation from the PWN FS at late times. In particular, external inverse-Compton scattering of jet photons by the FS naturally produces a late-time GeV bump together with a substantial TeV component. Our model can simultaneously account for early thermal emission, X-ray plateaus, and late-time GeV excesses in merger-driven gamma-ray bursts, and also indicates that post-merger magnetar-driven PWNs are potential TeV photon sources.
\end{abstract}

\keywords{Neutron star --- Jet --- Gamma-ray burst --- PWN }
	
\section{Introduction}\label{sec:intro}
Binary neutron star (BNS) mergers are key sources for multi-messenger astronomy, producing both gravitational-wave and electromagnetic signals (\citealp{1989Natur.340..126E,1998ApJ...507L..59L,2010MNRAS.406.2650M,2012ApJ...746...48M,2014ARA&A..52...43B}). This was exemplified by the landmark detection of GW170817 by the LIGO and Virgo collaborations in 2017 (\citealp{2017PhRvL.119p1101A}). The event yielded a clear gravitational-wave signal together with a short gamma-ray burst (GRB~170817A), kilonova emission in the optical and infrared bands, and long-lasting X-ray and radio afterglows (\citealp{2017ApJ...848L..12A,2017Natur.551...64A,2017Sci...358.1556C,2017ApJ...848L..14G}). Together, these observations marked the first comprehensive multi-messenger detection of a BNS merger. Some short GRBs of merger origin also exhibit long-timescale central-engine activity, such as extended emission and X-ray plateaus (\citealp{2006ApJ...643..266N,2013MNRAS.430.1061R,2014MNRAS.438..240G,2015JHEAp...7...73D,2020ApJ...895...58G,2020MNRAS.493..783M}). The recently observed GRB~211211A shows long-lasting prompt emission together with kilonova signatures, pointing to a merger origin (\citealp{2022Natur.612..228T,2022Natur.612..232Y,2022Natur.612..223R,2023NatAs...7...67G,2025ApJ...985..195L}). These long-timescale signals suggest that a long-lived central engine may remain after the merger, such as a rapidly rotating, highly magnetized neutron star (magnetar; \citealp{1998A&A...333L..87D,2001ApJ...552L..35Z,2011MNRAS.413.2031M,2018ApJ...869..155S,2024ApJ...970....6X}).

Depending on the equation of state and the binary masses, BNS mergers can leave different remnants, one of which is a millisecond magnetar (\citealp{2020GReGr..52..108B,2021GReGr..53...59S,2024ApJ...968..104L}). Such a magnetar can tap its rotational energy to power a relativistic jet. The jet is initially Poynting-flux dominated, and its magnetic energy can be converted into thermal radiation and bulk kinetic energy via gradual internal magnetic dissipation (\citealp{2001A&A...369..694S,2002A&A...387..714D,2002A&A...391.1141D,2005A&A...430....1G}). As the jet propagates outward, it interacts with the merger ejecta, driving forward and reverse shocks. The shocked region bounded by these shocks inflates a pulsar wind nebula (PWN; e.g., \citealp{1992ApJ...395..540C,2006ARA&A..44...17G}).

Magnetar-driven, Poynting-flux–dominated jets have been widely used to study GRB prompt emission (\citealp{2017MNRAS.472.3058B,2018ApJ...867...52X,2019ApJS..242...16L,2024A&A...691A.105M}). In addition, PWN emission has been investigated in the contexts of GRBs (\citealp{1998A&A...333L..87D,2001ApJ...552L..35Z,2007A&A...470..119Y,2018ApJ...855...67L}), superluminous supernovae (\citealp{2004ApJ...611..380T,2010ApJ...717..245K,2015ApJ...799..107W,2016ApJ...818...94K}), and kilonovae (\citealp{2015ApJ...807..163G,2017ApJ...837...50G,2019ApJ...885...60R,2025ApJ...978...52A}). However, most previous studies focus on only one component of this system, lacking a unified framework that self-consistently couples the magnetically dissipative jet, the shock dynamics of the surrounding nebula, and the resulting multi-wavelength---in particular very-high-energy---radiation in the post-merger phase.
In this study, we develop a self-consistent dynamical and radiative model that couples a magnetically dissipative Poynting jet to the PWN it inflates, and we compute the resulting multi-band emission from optical to TeV energies as the system evolves. Our goal is to obtain a comprehensive picture of the coupled jet--PWN dynamics and radiation at different evolutionary stages when a magnetar serves as the central engine.

This paper is organized as follows. In Section~\ref{sec2}, we present the dynamical framework and radiation processes for the Poynting-flux–dominated jet and the associated PWN. Section~\ref{sec3} shows the dynamical evolution, multi-wavelength light curves, and time-resolved spectra for representative parameter sets. Conclusions and discussion are given in Section~\ref{sec4}.

\section{Model} \label{sec2}

\subsection{Dynamics of the Poynting Jet} \label{sub2:1}

Newly born magnetars formed from neutron star mergers are typically highly magnetized and rapidly rotating, providing a vast reservoir of rotational energy with a total output up to $\sim 10^{52}$~erg (\citealp{1992ApJ...392L...9D,2013ApJ...771L..26G,2020GReGr..52..108B,2021GReGr..53...59S}).
Accordingly, we adopt a total spin-down energy of $E_{\rm sd,0}=10^{52}$~erg.
The spin-down luminosity of the magnetar can be expressed as
\begin{equation}
	L_{\rm sd} = L_{\rm sd,0} \left(1 + \frac{t}{T_{\rm sd}}\right)^{-2},
	\label{e2}
\end{equation}
where $L_{\rm sd,0}$ is the initial spin-down luminosity and $T_{\rm sd} = E_{\rm sd,0}/L_{\rm sd,0}$ denotes the characteristic spin-down timescale.
The spin-down energy released by the magnetar powers a Poynting-flux-dominated jet, as described in the following.

We consider a Poynting-flux-dominated jet characterized by a ``striped wind'' magnetic field configuration (\citealp{1990ApJ...349..538C, 2001A&A...369..694S}), in which the scale of the stripe, $\lambda$, is related to the rotation frequency of the central engine as $\lambda \sim \pi c / \Omega \sim 10^7~\mathrm{cm}$ (\citealp{2017MNRAS.468.3202B}). 
As the flow propagates outward, magnetic energy is dissipated through reconnection of oppositely oriented field lines. 
The reconnection rate is controlled by the inflow velocity of the plasma towards the current sheet, $v_{\rm rec} = \epsilon c$, where the dimensionless parameter $\epsilon$ typically lies in the range $0.1$--$0.25$ (\citealp{2005MNRAS.358..113L, 2015ApJ...806..167G, 2015PhRvL.114i5002L}). 
In our model calculations, $\lambda$ and $\epsilon$ appear only through the combination $\lambda/\epsilon$, for which we adopt $\lambda/\epsilon = 10^8~\mathrm{cm}$ (\citealp{2017MNRAS.468.3202B}).

To quantify the magnetic field component in the flow, the magnetization parameter $\sigma$ is introduced, i.e., the ratio of the Poynting flux $F_{\rm P}$ to the matter flux $F_{\rm m}$. 
As the jet expands, magnetic energy is gradually dissipated and converted into kinetic energy, causing $\sigma$ to decrease and the jet Lorentz factor $\Gamma_{\rm jet}$ to increase. 
The evolution of $\sigma$ can be expressed as 
$\sigma = \Gamma_{\rm sat}/\Gamma_{\rm jet}$,
where $\Gamma_{\rm sat} = \sigma_0^{3/2}$ is the maximum achievable Lorentz factor of the jet, i.e., 
the bulk Lorentz factor of the outflow at the saturation radius $R_{\rm sat} = \lambda \Gamma_{\rm sat}^2/(6\epsilon)$ (\citealp{2002A&A...391.1141D,2017MNRAS.468.3202B}). 
Here, $\sigma_0 = 100$ denotes the initial magnetization parameter (\citealp{2008A&A...480..305G}).
Following \citet{2002A&A...391.1141D}, the radial evolution of $\Gamma_{\rm jet}$ is given by
\begin{equation}
	\Gamma_{\rm jet} = 
	\begin{cases} 
		\Gamma_{\rm sat} \left( \dfrac{R}{R_{\rm sat}} \right)^{1/3}, & R_{\rm acc} < R \le R_{\rm sat}, \\[10pt]
		\Gamma_{\rm sat}, & R > R_{\rm sat},
	\end{cases}
\end{equation}
where $R_{\rm acc}=(\Gamma_{\rm jet0}/\Gamma_{\rm sat})^3 R_{\rm sat}$ denotes the jet acceleration radius. We take the initial jet Lorentz factor to be $\Gamma_{\rm jet0}=\sqrt{\sigma_0+1}$, which corresponds to the Alfv\'en speed (\citealp{2017MNRAS.468.3202B}).

\subsection{Evolution of the Pulsar Wind Nebula} \label{sub2:2}

As the jet propagates outward and interacts with the external medium, a pair of forward and reverse shocks (FS and RS) are generated, leading to the formation of a relativistic stellar wind bubble (see, e.g., Fig.~B1 in \citealp{2013MNRAS.433.2107N}). The Poynting-flux–dominated jet is separated from the external medium by a contact discontinuity (CD), and the entire system can be divided into four regions by the FS and RS: the unshocked ambient medium (Region~1), the forward-shocked ambient medium (Region~2), the reverse-shocked jet (Region~3), and the unshocked jet (Region~4). 
The dynamical evolution of the PWN can be determined from the evolution of its Lorentz factor $\Gamma$, internal energy $E'_{\rm int}$, swept-up mass $m$, and radius $R$ in Regions~2 and~3 (\citealp{2013MNRAS.433.2107N,2018pgrb.book.....Z}):
\begin{equation} \label{equ4}
	\begin{aligned}
		\frac{d\Gamma_{3}}{dR} = 
		- \frac{
			(\Gamma_{\rm eff} + 1)(\Gamma_{3} - 1)\frac{dm_2}{dR}c^2 
			+ (\Gamma_{3} - f_{\rm a} \Gamma_{\rm eff} + f_{\rm a} \Gamma_{\rm eff} \Gamma_{34})\frac{dm_{3}}{dR}c^2 
			+ \Gamma_{\rm eff} \frac{dE'_{\rm ad,3}}{dR} 
			+ \Gamma_{\rm eff} \frac{dE'_{\rm ad,2}}{dR} 
			- \frac{dE_{\rm sd}}{dR}
		}{
			(m_{2} + m_3)c^2 
			+ E'_{\rm int,2} \frac{\Gamma_{\rm eff}}{\Gamma_{3}} 
			+ E'_{\rm int,3} \frac{\Gamma_{\rm eff}}{\Gamma_{3}}
		}.
	\end{aligned}
\end{equation}
The internal energy of the shocked regions is expressed as
\begin{equation}
	E'_{\rm int,2}(R) = c^2 \int_0^R \left\{ dm_2(r_{\rm fs}) \mu_{\rm p} [\gamma_{\rm ad,p,2}(R,r) - 1] + dm_2(r_{\rm fs}) \mu_{\rm e} [\gamma_{\rm ad,e,2}(R,r) - 1] \right\},
\end{equation}
\begin{equation}
	E'_{\rm int,3}(R) = c^2 \int_0^R \left\{ dm_{3}(r_{\rm rs}) \mu_{\rm p} [\gamma_{\rm ad,p,3}(R,r) - 1] + dm_{3}(r_{\rm rs}) \mu_{\rm e} [\gamma_{\rm ad,e,3}(R,r) - 1] \right\}.
\end{equation}
The swept-up masses evolve as
\begin{equation}
	\frac{dm_{2}}{dR} = 4\pi R^2\rho,
\end{equation}
\begin{equation} \label{equ:mass_evolution}
	\frac{dm_3}{dR} = \frac{L_{\rm sd, inj}}{c^3 \Gamma_4 (1 + \sigma)} \frac{\beta_4 - \beta_{\rm rs}}{\beta_3\beta_4},
\end{equation}
and the radial evolution follows
\begin{equation} \label{equ:R_evolution}
	\frac{dR}{dt_{\rm obs}} = \frac{\beta_3 c}{1-\beta_3}.
\end{equation}

In our model, the prime symbol ($'$) denotes quantities measured in the comoving frame of the shocked medium. The subscripts “fs”, “rs” and “cd” refer to the FS, RS, and CD respectively, while the numbers “1–4” correspond to the four regions defined above. In these dynamical equations, the Lorentz factors of Regions~2 and~3 are equal, i.e., $\Gamma_{3} = \Gamma_{2}$. The initial Lorentz factor of Region~3, $\Gamma_{3,0} = 1.1$, is adopted. The effective Lorentz factor is defined as $\Gamma_{\rm eff} = (\hat{\gamma}\Gamma_3^2 - \hat{\gamma} + 1)/\Gamma_3$, where the adiabatic index is $\hat{\gamma} = (4 + \Gamma_3^{-1})/3$. The relative Lorentz factor between Regions~4 and~3 is $\Gamma_{34} = \Gamma_3 \Gamma_4 (1 - \beta_3 \beta_4)$, with $\Gamma_4 = \Gamma_{\rm jet}$ and $\beta = \sqrt{1 - 1/\Gamma^2}$ being the corresponding dimensionless velocity. 
The magnetization correction factor $f_{\rm a}$ is approximated as (\citealp{2005ApJ...628..315Z,2010A&A...518A..27M}):
\begin{equation}
	f_{\rm a} \approx 1 - \frac{\sigma}{2u_{\rm 3s}^2 + 2u_{\rm 3s}\sqrt{u_{\rm 3s}^2 + 1}},
\end{equation}
where $u_{\rm 3s}$ is the four-velocity of Region~3 measured in the rest frame comoving with the reverse shock. For $\Gamma_{34} \gg 1$, $u_{\rm 3s}$ can be analytically obtained from \citep{2004A&A...424..477F,2005ApJ...628..315Z}:
\begin{equation}
	8(\sigma + 1) u_{\rm 3s}^4 - (8\sigma^2 + 10\sigma + 1) u_{\rm 3s}^2 + \sigma^2 = 0.
\end{equation}

The adiabatic losses in Region~2 can be expressed as \citep{2013MNRAS.433.2107N}:
\begin{equation}
	\frac{dE'_{\rm ad,2}(R)}{dR} = -c^2 \left( \frac{1}{R} - \frac{1}{3} \frac{d\log \Gamma_3}{dR} \right) \int_0^R \left[ dm(r_{\rm fs}) \mu_{\rm p} \frac{p_{\rm ad,p,2}^2(R,r)}{\gamma_{\rm ad,p,2}(R,r)} + dm(r_{\rm fs}) \mu_{\rm e} \frac{p_{\rm ad,e,2}^2(R,r)}{\gamma_{\rm ad,e,2}(R,r)} \right].
\end{equation}
Due to adiabatic expansion, the Lorentz factors of post-shock protons and electrons decrease with time. Their instantaneous values at radius $R$ are given by $\gamma_{\rm ad,p,2} = \sqrt{p_{\rm ad,p,2}^2 + 1}$ and $\gamma_{\rm ad,e,2} = \sqrt{p_{\rm ad,e,2}^2 + 1}$, where $p_{\rm ad,p,2}$ and $p_{\rm ad,e,2}$ denote the comoving momenta. For a particle injected at radius $r$ with initial momentum $p(r)$, the comoving momenta evolve as
\begin{equation}
	p_{\rm ad,p,2}(R,r) = \frac{r_{\rm fs}}{r_{\rm fs} + R_{\rm cd} - r_{\rm cd}} \left[ \frac{\Gamma(R)}{\Gamma(r)} \right]^{1/3} \left[ \gamma_{\rm acc,p,2}^2(r) - 1 \right]^{1/2},
\end{equation}
\begin{equation}
	p_{\rm ad,e,2}(R,r) = \frac{r_{\rm fs}}{r_{\rm fs} + R_{\rm cd} - r_{\rm cd}} \left[ \frac{\Gamma(R)}{\Gamma(r)} \right]^{1/3} \left[ \gamma_{\rm rad,e,2}^2(r) - 1 \right]^{1/2}.
\end{equation}
The mean Lorentz factors of freshly shocked protons and electrons at the injection radius $r$ are
\begin{equation}
	\gamma_{\rm acc,p,2}(r) = [\Gamma(r) - 1] \frac{\epsilon_{\rm p}}{\mu_{\rm p}} + 1,
\end{equation}
\begin{equation}
	\gamma_{\rm rad,e,2}(r) = (1 - \epsilon_{\rm rad}) [\Gamma(r) - 1] \frac{\epsilon_{\rm e}}{\mu_{\rm e}} + 1,
\end{equation}
where $\epsilon_{\rm rad}$ denotes the fraction of electron energy lost to radiation, $\epsilon_{\rm e}\sim 0.1$ and $\epsilon_{\rm p}=1-\epsilon_{\rm e}$ represent the fractions of total dissipated energy transferred to protons and electrons, respectively. The parameters $\mu_{\rm p} \simeq 1$ and $\mu_{\rm e} \simeq m_{\rm e}/m_{\rm p}$ denote the mass fractions of shocked protons and electrons, respectively. 
The expressions for adiabatic losses in Region~3 are analogous to those in Region~2, except that the mean Lorentz factors of freshly shocked particles are modified by the magnetization correction factor:
\begin{equation}
	\gamma_{\rm acc,p,3} = f_{\rm a} (\Gamma_{34} - 1) \frac{\epsilon_{\rm p}}{\mu_{\rm p}} + 1,
\end{equation}
\begin{equation}
	\gamma_{\rm rad,e,3} = f_{\rm a} (1 - \epsilon_{\rm rad}) (\Gamma_{34} - 1) \frac{\epsilon_{\rm e}}{\mu_{\rm e}} + 1.
\end{equation}

Suppose that the unshocked ejecta has a wind-like density profile and that the forward shock eventually propagates into the outer interstellar medium (ISM), i.e.,
\begin{equation}
	\rho(R) = \left\{ 
	\begin{array}{ll}
		A_{0}R^{-2}, & \quad R_{\rm ej,0}< R \leq R_{\rm fin}, \\[10pt]
		n_{\rm ISM}m_{\rm p}, & \quad R > R_{\rm fin},
	\end{array} 
	\right.
\end{equation}
where the initial radius of the ejecta is $R_{\rm ej,0}=5\times 10^{10}\,\mathrm{cm}$ and the outer radius is taken as $R_{\rm fin}=100\,R_{\rm ej,0}$. Here $n_{\rm ISM}$ is the ISM number density (we adopt a fiducial value $n_{\rm ISM}=0.1~\mathrm{cm^{-3}}$), and $A_{0}$ is the normalization constant determined by
\begin{equation}
	\int_{R_{\rm ej,0}}^{R_{\rm fin}}4\pi R^2\rho(R)\,dR = M_{\rm ej}.
\end{equation}
where $M_{\rm ej}$ is the mass of the ejecta.
Additionally, the energy injected into Region~3 by the magnetar’s spin-down is given by
\begin{equation} \label{equ:spin_down_energy E}
	\frac{dE_{\rm sd}}{dt} = L_{\rm sd,inj} \frac{\beta_4 - \beta_{\rm rs}}{\beta_4},
\end{equation}
where the injection luminosity is
\begin{equation}
	L_{\text{sd,inj}} = 
	\begin{cases}
		\xi_{\rm rs} L_{\text{sd}}(t_{\rm inj}), & t_{\text{inj}} \ge 0, \\[6pt]
		0, & t_{\text{inj}} < 0,
	\end{cases}
\end{equation}
where $t_{\rm inj}=(c\beta_{\rm jet}t+R_{\rm acc}-R_{\rm rs})/(c\beta_{\rm jet})$ is the injection time in the source frame, and $\xi_{\rm rs}$ parameterizes the fraction of the spin-down power injected into the shock. We adopt $\xi_{\rm rs}=0.5$ as a representative value, similar to that used in previous post-merger magnetar wind calculations, while emphasizing that the exact efficiency remains uncertain (\citealp{2013ApJ...771...86G,2016A&A...592A..92L}).
Finally, the evolution of the forward shock (FS), reverse shock (RS), and contact discontinuity (CD) radii is
\begin{equation}
	\frac{dR_{\rm fs}}{dt} = \beta_{\rm fs} c, \quad 
	\frac{dR_{\rm rs}}{dt} = \beta_{\rm rs} c, \quad 
	\frac{dR_{\rm cd}}{dt} = \beta_3 c,
\end{equation}
where the shock velocities are determined by the shock jump conditions (\citealp{2005ApJ...628..315Z,2017ApJ...834...32S}):
\begin{equation}
	\beta_{\rm fs} = \frac{ \hat{\gamma} \Gamma_2^2 (-\beta_2) \beta_2 - (\hat{\gamma} - 1)(1 - \Gamma_2) }{ \hat{\gamma} \Gamma_2^2 (-\beta_2) - (\hat{\gamma} - 1)(-\Gamma_2 \beta_2) },
\end{equation}
\begin{equation}
	\beta_{\rm rs} = \frac{ \beta_{3} - \beta_{\rm 3s} }{ 1 - \beta_{3}\beta_{\rm 3s} },
\end{equation}
where $\beta_{3s} = u_{\rm 3s}/\sqrt{1+u_{\rm 3s}^2}$.

\subsection{Radiation from the Jet} \label{sub2:3}

As the jet dissipates its energy, a fraction is converted into radiation. 
The radiation component consists of both thermal emission produced via internal gradual magnetic dissipation and non-thermal synchrotron radiation generated by electrons accelerated through magnetic reconnection (\citealp{2001A&A...369..694S,2002A&A...387..714D,2002A&A...391.1141D,2014MNRAS.445.3892B,2014ApJ...783L..21S,2015SSRv..191..545K}). 
Throughout this work, we adopt the convention $Q = 10^x Q_x$ in cgs units.

At a given radius, the Poynting-flux luminosity could be written as (\citealp{2005A&A...430....1G,2017MNRAS.468.3202B})
\begin{equation}
	L_{\rm B} = c\frac{(RB)^2}{4\pi} = L_{\rm sd}\left(1 - \frac{\Gamma_{\rm jet}}{\Gamma_{\rm sat}}\right),
	\label{e25}
\end{equation}
where $B$ is the magnetic field strength at radius $R$. 
As the co-moving temperature decreases with radius as $T' \propto R^{-7/9}$, the thermal luminosity evolves as $L_{\rm th} \propto R^{-4/9}$ (\citealp{2005A&A...430....1G}). 
Substituting the energy dissipation rate $d\dot{E} = - (dL_{\rm B}/dR) dR$, the total thermal photospheric luminosity can be estimated as (\citealp{2005A&A...430....1G,2017MNRAS.468.3202B,2017ApJ...846..130X})
\begin{equation}
	L_{\rm ph} \propto \int_0^{R_{\rm ph}} \frac{dR}{R^{2/3}} \left( \frac{R}{R_{\rm ph}} \right)^{4/9}
	= 6.6\times10^{50} L_{\rm sd,52}^{6/5} \left( \frac{\lambda}{\epsilon} \right)_8^{-1/5} 
	\Gamma_{\rm sat,3}^{-1}~\rm erg\,s^{-1}\,sr^{-1}.
\end{equation}
The corresponding photospheric temperature is
\begin{equation}
	T_{\rm ph} = 110\, L_{\rm sd,52}^{3/10} \Gamma_{\rm sat,3}^{1/4} 
	\left( \frac{\lambda}{\epsilon} \right)_8^{-11/20}~\rm keV,
\end{equation}
and the photospheric radius is given by (\citealp{2017MNRAS.468.3202B,2017ApJ...846..130X})
\begin{equation}
	R_{\rm ph} = 4.6\times10^{11} 
	\left( \frac{\lambda}{\epsilon} \right)_8^{2/5}
	\frac{L_{\rm sd,52}^{3/5}}{\Gamma_{\rm sat,3}}~\rm cm.
\end{equation}
Thus, the thermal component can be expressed as
\begin{equation}
	L_{\nu}^{\rm ph}(\nu) = L_{\nu}^{\rm ph}(\nu; L_{\rm sd}, \lambda/\epsilon, \Gamma_{\rm sat}).
\end{equation}

For the non-thermal component, we calculate the characteristic synchrotron frequencies: 
$\nu_{\rm m}$ (the minimum synchrotron frequency), 
$\nu_{\rm c}$ (the cooling frequency), and 
$\nu_{\rm max}$ (the maximum synchrotron frequency), following \cite{2017MNRAS.468.3202B} and \cite{2017ApJ...846..130X}:
\begin{align}
	\nu_{\rm m} &= \Gamma_{\rm jet} \gamma_{\rm m}^2 \frac{q B'}{2\pi m_e c}
	\approx 1.4 \times 10^{19} 
	\left( \frac{R_{\rm sat}}{R} \right)^{5/3}
	\frac{L_{\rm sd,52}^{1/2}}{\Gamma_{\rm sat,3}^2}
	\left( \frac{\epsilon_{\rm e,jet}}{\xi} \right)^2
	\left( \frac{\lambda}{\epsilon} \right)_8^{-1}~\rm Hz, \\[4pt]
	\nu_{\rm c} &= 
	\frac{72\pi q m_e c^3 \Gamma_{\rm jet}^3}{\sigma_{\rm T}^2 B'^3 R^2}
	\approx 10^{16}
	\left( \frac{R}{R_{\rm sat}} \right)^3
	\frac{\Gamma_{\rm sat,3}^8}{L_{\rm sd,52}^{3/2}}
	\left( \frac{\lambda}{\epsilon} \right)_8~\rm Hz, \\[4pt]
	\nu_{\rm max} &= \Gamma_{\rm jet} \gamma_{\rm max}^2 
	\frac{q B'}{2\pi m_e c}.
\end{align}

Here, $q$ is the electron charge, $m_e$ is the electron mass, $\sigma_{\rm T}$ is the Thomson cross-section, 
$\epsilon_{\rm e,jet} \sim 0.2$ is the fraction of dissipated energy per electron, 
and $\xi \simeq 0.2$ is the fraction of electrons accelerated in the reconnection regions (\citealp{2015MNRAS.450..183S}). 
Combining Equation~(\ref{e2}) with Equation~(\ref{e25}), the co-moving magnetic field strength of the jet can be expressed as (\citealp{2017MNRAS.468.3202B})
\begin{equation}
	B' = \frac{B}{\Gamma_{\rm jet}} 
	= \frac{4.1\times10^6}{R_{12}^{4/3}}
	\frac{L_{\rm sd,52}^{1/2}}{\Gamma_{\rm sat,3}^{1/3}}
	\left( \frac{\lambda}{\epsilon} \right)_8^{1/3}~\rm G.
\end{equation}
The minimum Lorentz factor $\gamma_{\rm m}$ depends on the electron energy spectrum. 
PIC simulations suggest that the accelerated electrons follow a power-law distribution with index $p$, where $p = 4\sigma^{-0.3}$ (\citealp{2014ApJ...783L..21S,2015ApJ...806..167G,2015SSRv..191..545K,2016ApJ...816L...8W}). 
For $p < 2$, we adopt $\gamma_{\rm m} \simeq 1$. 
For $p > 2$, it can be written as (\citealp{2017MNRAS.468.3202B})
\begin{equation}
	\gamma_{\rm m} = \frac{p - 2}{p - 1}
	\frac{\epsilon_{\rm e,jet}}{2\xi}
	\sigma \frac{m_{\rm p}}{m_{\rm e}}.
\end{equation}
The maximum Lorentz factor is given by (\citealp{2010MNRAS.408L..46G,2017ApJ...846..130X})
\begin{equation}
	\gamma_{\rm max} = 
	\left( \frac{6\pi q \epsilon}{\sigma_{\rm T} B'} \right)^{1/2}.
\end{equation}

Depending on the relative ordering of the characteristic frequencies, the electrons may be in either the fast- or slow-cooling regime. In our model, these characteristic frequencies depend on the adopted dynamical and radiative parameters. For the fiducial parameter sets considered here, electrons at the initial photospheric radius $R_{\rm ph}$ are initially in the fast-cooling regime. Specifically, for $L_{\rm sd,0}=10^{49}\,\mathrm{erg\,s^{-1}}$ and $10^{50}\,\mathrm{erg\,s^{-1}}$, we obtain $\nu_m = 1.93\times10^{16}\,\mathrm{Hz}$ and $9.63\times10^{21}\,\mathrm{Hz}$, respectively, both of which are significantly higher than the corresponding cooling frequencies, $\nu_c = 2.50\times10^{10}\,\mathrm{Hz}$ and $4.98\times10^{10}\,\mathrm{Hz}$. This ordering, $\nu_m>\nu_c$, indicates that the electrons are initially in the fast-cooling regime for these representative cases. Neglecting synchrotron self-absorption, the spectrum in the fast-cooling regime, i.e., when $\nu_m>\nu_c$, is (\citealp{1998ApJ...497L..17S})
\begin{equation}
	L_{\nu}^{\mathrm{syn}} =
	\begin{cases}
		L_{\nu,\max}^{\mathrm{syn}} \left( \nu / \nu_c \right)^{1/3}, & \text{if } \nu < \nu_c, \\[6pt]
		L_{\nu,\max}^{\mathrm{syn}} \left( \nu / \nu_c \right)^{-1/2}, & \text{if } \nu_c < \nu < \nu_m, \\[6pt]
		L_{\nu,\max}^{\mathrm{syn}} \left( \nu_m / \nu_c \right)^{-1/2}
		\left( \nu / \nu_m \right)^{-p/2}, & \text{if } \nu_m < \nu < \nu_{\max}.
	\end{cases}
\end{equation}
with 
\begin{equation}
	L_{\nu,\max}^{\mathrm{syn}} = 
	\frac{m_e c^2 \sigma_{\rm T} \Gamma_{\rm jet} B' N_e(R)}{3q}=5\times10^{34}	\frac{L_{\rm sd,52}^{1.3}}{\Gamma_{\rm sat,3}^2}
	\left( \frac{\xi}{0.2} \right)
	\left( \frac{\lambda}{\epsilon} \right)_8^{1/5}~ \rm erg \, Hz^{-1} \, s^{-1},
\end{equation}
where $N_e(R)$ is the total number of radiating electrons at radius $R$.
 
In the fiducial cases considered here, the ratio $\nu_m/\nu_c$ decreases with radius. The transition from the fast- to the slow-cooling regime occurs at the characteristic radius $R_{\rm tr}$, defined by the condition
$\nu_m=\nu_c$. For $R_{\rm tr}<R<R_{\rm sat}$, the electron population enters the
slow-cooling regime, where $\nu_m<\nu_c$. In the absence of synchrotron self-absorption, the corresponding slow-cooling spectrum is given by  (\citealp{1998ApJ...497L..17S})

\begin{equation}
	L_{\nu}^{\rm syn} =
	\begin{cases}
		L_{\nu,\max}^{\rm syn}
		\left( \nu / \nu_m \right)^{1/3},
		& \nu < \nu_m, \\[6pt]
		L_{\nu,\max}^{\rm syn}
		\left( \nu / \nu_m \right)^{-(p-1)/2},
		& \nu_m < \nu < \nu_c, \\[6pt]
		L_{\nu,\max}^{\rm syn}
		\left( \nu_c / \nu_m \right)^{-(p-1)/2}
		\left( \nu / \nu_c \right)^{-p/2},
		& \nu_c < \nu < \nu_{\max}.
	\end{cases}
\end{equation}

Synchrotron self-absorption may modify the low-frequency part of the spectrum. The self-absorption frequency $\nu_a$ and the corresponding
electron Lorentz factor $\gamma_a$ are determined by
\begin{equation}
	\frac{2\nu_a^2}{c^2}
	\gamma_a \Gamma_{\rm jet} m_e c^2
	\frac{\pi R^2}{\Gamma_{\rm jet}^2}
	= L_{\nu_a}^{\rm syn},
\end{equation}
with
\begin{equation}
	\nu_a
	= \Gamma_{\rm jet} \gamma_a^2
	\frac{qB'}{2\pi m_e c}.
\end{equation}
At the initial photospheric radius $R_{\rm ph}$, the self-absorption frequency is larger than the cooling frequency, i.e., $\nu_a>\nu_c$. As the jet expands, $\nu_a$ crosses	$\nu_c$ at a characteristic radius $R_{\rm cr}$. 
Following \cite{2017ApJ...846..130X}, the full synchrotron spectrum can be written as:

For $R_{\rm ph} < R \leq R_{\rm cr}$,
\begin{equation}
	L_{\nu}^{\mathrm{syn}} =
	\begin{cases}
		L_{\nu_a}^{\mathrm{syn}} \left( \nu / \nu_a \right)^{11/8}, & \text{if } \nu < \nu_a, \\[6pt]
		L_{\nu_a}^{\mathrm{syn}} \left( \nu / \nu_a \right)^{-1/2}, & \text{if } \nu_a < \nu < \nu_m, \\[6pt]
		L_{\nu_a}^{\mathrm{syn}} \left( \nu_m / \nu_a \right)^{-1/2} 
		\left( \nu / \nu_m \right)^{-p/2}, & \text{if } \nu_m < \nu < \nu_{\max}.
	\end{cases}
\end{equation}
For $R_{\rm cr} < R \leq R_{\rm tr}$,
\begin{equation}
	L_{\nu}^{\mathrm{syn}} =
	\begin{cases}
		L_{\nu_a}^{\mathrm{syn}} \left( \nu / \nu_a \right)^{11/8}, & \text{if } \nu < \nu_a, \\[6pt]
		L_{\nu_a}^{\mathrm{syn}} \left( \nu / \nu_a \right)^{1/3}, & \text{if } \nu_a < \nu < \nu_c, \\[6pt]
		L_{\nu_a}^{\mathrm{syn}} \left( \nu_c / \nu_a \right)^{1/3} 
		\left( \nu / \nu_c \right)^{-1/2}, & \text{if } \nu_c < \nu < \nu_m, \\[6pt]
		L_{\nu_a}^{\mathrm{syn}} \left( \nu_c / \nu_a \right)^{1/3} 
		\left( \nu_m / \nu_c \right)^{-1/2}
		\left( \nu / \nu_m \right)^{-p/2}, & \text{if } \nu_m < \nu < \nu_{\max}.
	\end{cases}
\end{equation}
Finally, for $R_{\rm tr}<R\leq R_{\rm sat}$, 
\begin{equation}
	L_{\nu}^{\mathrm{syn}} =
	\begin{cases}
		L_{\nu_a}^{\mathrm{syn}} \left( \nu / \nu_a \right)^{11/8}, & \text{if } \nu < \nu_a, \\[6pt]
		L_{\nu_a}^{\mathrm{syn}} \left( \nu / \nu_a \right)^{1/3}, & \text{if } \nu_a < \nu < \nu_m, \\[6pt]
		L_{\nu_a}^{\mathrm{syn}} \left( \nu_m / \nu_a \right)^{1/3} 
		\left( \nu / \nu_m \right)^{-(p-1)/2}, & \text{if } \nu_m < \nu < \nu_c, \\[6pt]
		L_{\nu_a}^{\mathrm{syn}} \left( \nu_m / \nu_a \right)^{1/3} 
		\left( \nu_c / \nu_m \right)^{-(p-1)/2} 
		\left( \nu / \nu_c \right)^{-p/2}, & \text{if } \nu_c < \nu < \nu_{\max}.
	\end{cases}
\end{equation}

The total non-thermal synchrotron emission is obtained by integrating the above
expressions from $R_{\rm ph}$ to $R_{\rm sat}$, yielding
\begin{equation}
	L_{\nu}^{\mathrm{syn}}(\nu)
	= L_{\nu}^{\mathrm{syn}}
	\bigl(\nu; L_{\rm sd}, \lambda/\epsilon, \epsilon_{\rm e,jet}, \Gamma_{\rm sat}, p \bigr).
\end{equation}
The total jet luminosity, consisting of both the thermal and non-thermal
components, can then be written as
\begin{equation}
	L_{\nu,\rm jet}(\nu)
	=
	L_{\nu}^{\rm ph}
	\bigl(\nu; L_{\rm sd}, \lambda/\epsilon, \Gamma_{\rm sat}\bigr)
	+
	L_{\nu}^{\mathrm{syn}}
	\bigl(\nu; L_{\rm sd}, \lambda/\epsilon, \epsilon_{\rm e,jet}, \Gamma_{\rm sat}, p \bigr).
\end{equation}
Accordingly, the observed flux is
\begin{equation}
	F_{\nu}(\nu)
	=\frac{L_{\nu,\rm jet}(\nu)}{4\pi D_{L}^{2}},
\end{equation}
where $D_L$ is the luminosity distance.

\subsection{Emission Processes in PWN} \label{sub2:4}
\subsubsection{Electron Spectrum in the Shocked Regions}\label{sub2:4.1}
After the Poynting–flux–dominated jet interacts with the external medium, a pair of FS and RS are formed. As they propagate, these shocks sweep up and accelerate electrons, leading to non-thermal emission. We assume that the injection rates of the accelerated electrons follow power-law distributions, i.e.,
$Q_{\rm fs}' = K_{\rm fs}(\gamma_{\rm e}'/\gamma_{\rm m,fs}')^{-p_{\rm fs}}$ and
$Q_{\rm rs}' = K_{\rm rs}(\gamma_{\rm e}'/\gamma_{\rm m,rs}')^{-p_{\rm rs}}$, respectively,
and adopt $p_{\rm fs}=p_{\rm rs}=2.3$. The normalization constants $K_{\rm fs}$ and $K_{\rm rs}$ are determined by
\begin{equation}
	\int_{\gamma_{\rm m,fs}'}^{\gamma_{\rm max,fs}'} Q_{\rm fs}'\,{\rm d}\gamma_{\rm e}'
	\;=\; 4\pi R_{\rm fs}^{2}\, n_{1}'\, \beta_{\rm fs} c,
\end{equation}
\begin{equation}
	\int_{\gamma_{\rm m,rs}'}^{\gamma_{\rm max,rs}'} Q_{\rm rs}'\,{\rm d}\gamma_{\rm e}'
	\;=\; 4\pi R_{\rm rs}^{2}\, n_{4}'\,
	\frac{\beta_{4}-\beta_{\rm rs}}{1-\beta_{4}\beta_{\rm rs}}\; c,
\end{equation}
where primes ($'$) denote quantities measured in the comoving frame of the shocked medium. For $\gamma_{\rm max,fs}' \gg \gamma_{\rm m,fs}'$ and $\gamma_{\rm max,rs}' \gg \gamma_{\rm m,rs}'$, the explicit forms of the injection rates are
\begin{equation}
	Q_{\rm fs}' \;=\; 4\pi R_{\rm fs}^{2}\, n_{1}'\,(p_{\rm fs}-1)\,
	\gamma_{\rm m,fs}'^{\,p_{\rm fs}-1}\, \gamma_{\rm e}'^{-p_{\rm fs}}\,
	\beta_{\rm fs} c,
\end{equation}
\begin{equation}
	Q_{\rm rs}' \;=\; 4\pi R_{\rm rs}^{2}\, n_{4}'\,(p_{\rm rs}-1)\,
	\gamma_{\rm m,rs}'^{\,p_{\rm rs}-1}\, \gamma_{\rm e}'^{-p_{\rm rs}}\,
	\frac{\beta_{4}-\beta_{\rm rs}}{1-\beta_{4}\beta_{\rm rs}}\; c,
\end{equation}
with $Q_{\rm fs}'=0$ and $Q_{\rm rs}'=0$ outside their respective energy ranges
($\gamma_{\rm m,fs}' \le \gamma_{\rm e}' \le \gamma_{\rm max,fs}'$ and
$\gamma_{\rm m,rs}' \le \gamma_{\rm e}' \le \gamma_{\rm max,rs}'$).
The electron number densities in Regions 1 and 4 in their respective comoving frames are given by
$n_{1}'=\Gamma_{\rm fs} n_{1}$ and
(\citealp{2016A&A...592A..92L})
\begin{equation}
	n_{4}' \;=\; \frac{L_{\rm sd,inj}}{4\pi R_{\rm rs}^{2}\, \Gamma_{4}^{2}\, m_{\rm e} c^{3}\,(1+\sigma)}.
\end{equation}

The minimum and maximum electron Lorentz factors are (\citealp{2016A&A...592A..92L})
\begin{equation}
	\gamma_{\rm m,fs}' \;=\; \epsilon_{\rm e,fs}\!\left(\frac{p_{\rm fs}-2}{p_{\rm fs}-1}\right)\!
	\frac{m_{\rm p}}{m_{\rm e}}\; \Gamma_{\rm fs},
\end{equation}
\begin{equation}
	\gamma_{\rm m,rs}' \;=\; \epsilon_{\rm e,rs}\!\left(\frac{p_{\rm rs}-2}{p_{\rm rs}-1}\right)\!
	(\Gamma_{34}-1)\, f_{\rm a},
\end{equation}
and
\begin{equation}
	\gamma_{\rm max,fs}' \;=\; \sqrt{\frac{3 q_{\rm e} B_{\rm fs}'}{4 \sigma_{\rm T}\,(U_{\rm fs,B}'+U_{\rm fs,eic}')}}, 
\end{equation}
\begin{equation}
	\gamma_{\rm max,rs}' \;=\; \sqrt{\frac{3 q_{\rm e} B_{\rm rs, tot}'}{4 \sigma_{\rm T}\,(U_{\rm rs,B}'+U_{\rm rs,eic}')}}.
\end{equation}
Here $B_{\rm fs}' \;=\; \sqrt{32\pi\, \epsilon_{\rm B,fs}\, n_{1}'\, m_{\rm p} c^{2}\, (\Gamma_{\rm fs}-1)}$,
and the total comoving magnetic field in the RS region is the quadrature sum of the advected and shock-generated components,
\begin{equation}
	B_{\rm rs, tot}' \;=\; \sqrt{B_{3}'^{2}+B_{\rm rs}'^{2}},
\end{equation}
with $B_{3}' \;=\; B_{4}'\,(4\Gamma_{34}+3)\, f_{\rm b}$, 
$B_{\rm rs}' \;=\; \sqrt{4\pi\, \epsilon_{\rm B,rs}\, (\Gamma_{34}-1)\, f_{\rm a}\, n_{3}'\, m_{\rm e} c^{2}}$,
and $B_{4}'=\sqrt{4\pi n_{4}' m_{\rm e} c^{2}\, \sigma}$ the comoving magnetic field in Region~4 (\citealp{2016A&A...592A..92L}). The correction factor $f_{\rm b}$ is (\citealp{2005ApJ...628..315Z,2010A&A...518A..27M})
\begin{equation}
	f_{\rm b} \;=\; \frac{1}{4}\!\left[\,1+\left(1+\frac{1}{u_{\rm 3s}^{2}}\right)^{1/2}\right].
\end{equation}
The comoving magnetic energy densities are
$U_{\rm fs,B}'=B_{\rm fs}'^{2}/8\pi$ and $U_{\rm rs,B}'=B_{\rm rs,tot}'^{2}/8\pi$,
while $U_{\rm fs,eic}'$ and $U_{\rm rs,eic}'$ represent the seed-photon energy densities for external inverse Compton (EIC) scattering in the FS and RS regions, respectively. 
Unless otherwise stated, we adopt the energy-partition fractions
	$\epsilon_{\rm e,fs}=\epsilon_{\rm e,rs}=0.1$,
	$\epsilon_{\rm B,fs}=10^{-4}$,
	and $\epsilon_{\rm B,rs}=0.1$,
	motivated by typical GRB afterglow modeling, which generally favors
	$\epsilon_{\rm e}\sim 0.1$ and relatively low $\epsilon_{\rm B}$ in the forward shock,
	while allowing for a more strongly magnetized reverse-shock region
(\citealp{2005ApJ...628..315Z,2014ApJ...785...29S,2017MNRAS.472.3161B}).

Then, the electron spectrum $dN_{\rm e,rs}/d\gamma_{\rm e}'$ in the reverse-shocked region at radius $R$ can be obtained by solving the continuity equation of electrons \citep{2008MNRAS.384.1483F}:
\begin{equation}
	\frac{dR}{dt'} \frac{\partial}{\partial R} \left( \frac{dN_{\mathrm{e,rs}}}{d\gamma'_{\rm e}} \right)
	+ \frac{\partial}{\partial \gamma'_{\rm e}} \left( \dot{\gamma}'_{\mathrm{e,rs}} \frac{dN_{\mathrm{e,rs}}}{d\gamma'_{\rm e}} \right)
	= Q_{\mathrm{rs}}',
\end{equation}
where $dR/dt'=\Gamma_{3}\beta_{3}c$, and
$\dot{\gamma}'_{\mathrm{e,rs}} 
= \dot{\gamma}'^{\mathrm{syn}}_{\mathrm{e,rs}}
+ \dot{\gamma}'^{\mathrm{ad}}_{\mathrm{e,rs}}
+ \dot{\gamma}'^{\mathrm{ic}}_{\mathrm{e,rs}}$
is the total cooling rate of electrons with Lorentz factor $\gamma'_{\rm e}$.  
Here, $\dot{\gamma}'^{\mathrm{syn}}_{\mathrm{e,rs}}=-\sigma_{\rm T}B_{\rm rs,tot}'^{2}\gamma'^2_{\rm e}/(6\pi m_{\rm e}c)$ represents the synchrotron cooling rate,  
$\dot{\gamma}'^{\mathrm{ad}}_{\mathrm{e,rs}}=-2\gamma'_{\rm e}\Gamma_{\rm rs}\beta_{\rm rs}c/(3R)$ corresponds to adiabatic expansion cooling,  
and $\dot{\gamma}'^{\mathrm{ic}}_{\mathrm{e,rs}}$ (described in Equation~(\ref{IC cool})) corresponds to the inverse-Compton cooling rate.  
By replacing the subscript “rs” with “fs” in the above parameters and equations, one can obtain the corresponding expressions for the forward-shocked region.

\subsubsection{Thermal, Synchrotron and Inverse-Compton Emission}\label{sub2:4.2}

Initially, the ejecta is optically thick, and its comoving temperature is
\begin{equation}
	T' = \left[\frac{E'_{\rm int,2} + E'_{\rm int,3}}{aV'}\right]^{1/4},
\end{equation}
where the comoving volume is $V' = 4\pi R_{\rm cd}^{3}/3$, and $a$ is the radiation constant.  
For an observed frequency $\nu_{\rm obs}$, the observed thermal flux from the ejecta is given by \citep{2013ApJ...776L..40Y}
\begin{equation}
	F_{\rm th}(t_{\rm obs}) 
	= \frac{1}{2D_{\rm L}^2}\int_0^{\theta_{\rm jet}} 
	\frac{1}{\max(\tau, 1)} 
	\cdot \frac{8\pi^{2} \mathcal{D}^{2} R_{\rm cd}^{2}}{h^{3} c^{2}} 
	\cdot \frac{(h\nu')^{4}}{\exp\!\left[h\nu'/(kT')\right] - 1}
	\sin\theta \, \mathrm{d}\theta,
\end{equation}
where $\mathcal{D} = [\Gamma_{3}(1 - \beta_{3}\cos\theta)]^{-1}$ is the Doppler factor and $\nu' = (1+z)\nu_{\rm obs}/\mathcal{D}$. We adopt a redshift $z = 0.076$, similar to GRB~211211A (\citealp{2022Natur.612..223R,2022Natur.612..228T,2022Natur.612..232Y}). The jet half-opening angle is taken to be $\theta_{\rm jet} = 0.1~\rm rad$.
The corresponding observer time of photons emitted at radius $R = R_{\rm obs}$ and latitude $\theta$ is
$
	t_{\rm obs}
	= (1+z)\int_{R_{\rm ejecta,0}}^{R_{\rm obs}} 
	(1 - \beta_3 \cos\theta)/(\beta_3 c)\,\mathrm{d}R
	+ R_{\rm ejecta,0}/(2 \Gamma_{3}^2 c)
$.
The optical depth of the ejecta is
$
	\tau = \kappa (M_{\rm ej}/V')(R_{\rm cd}/\Gamma_{3}),
$
where $\kappa$ is the opacity. Following \citet{2019ApJ...877L..21Y}, for $h\nu \gtrsim 0.1~{\rm keV}$ we adopt
\begin{equation}
	\kappa =
	C_{1}\!\left( \frac{h\nu}{1~\mathrm{keV}} \right)^{-3}
	+ C_{2}\!\left[
	\left( \frac{h\nu}{50~\mathrm{keV}} \right)^{-0.36}
	+ \left( \frac{h\nu}{50~\mathrm{keV}} \right)^{0.65}
	\right]^{-1}
	+ C_{3},
\end{equation}
with $C_{1} = 2.0~{\rm cm^{2}\,g^{-1}}$, $C_{2} = 0.7~{\rm cm^{2}\,g^{-1}}$, and $C_{3} = 0.012~{\rm cm^{2}\,g^{-1}}$. For $h\nu < 0.1~{\rm keV}$, we set $\kappa = 0.2~{\rm cm^{2}\,g^{-1}}$ (\citealp{2013ApJ...776L..40Y}). Once the ejecta becomes optically thin ($\tau \lesssim 1$), its thermal emission rapidly becomes negligible. In our calculation, we simply assume that the blackbody radiation from the ejecta persists until the optical depth decreases to $\tau = 1$.

After the ejecta becomes optically thin, the non-thermal radiation from the PWN starts to dominate. The relevant non-thermal processes include synchrotron radiation and inverse-Compton (IC) scattering in both the forward- and reverse-shocked regions.
The synchrotron power per unit solid angle in the comoving frame of the reverse-shocked region at frequency $\nu'$ is
\begin{equation}
	P'_{\rm syn,rs}(\nu', R) = \frac{1}{4\pi} \int p'_{\rm syn,rs}(\nu', \gamma'_e) 
	\frac{\mathrm{d}N_{e,\rm rs}}{\mathrm{d}\gamma'_e} \, \mathrm{d}\gamma'_e,
\end{equation}
where the single-electron synchrotron power is
$	p'_{\rm syn,rs}(\nu', \gamma'_e) 
	= \sqrt{3}\, q_e^3 B'_{\rm rs,tot}/(m_e c^2)F(\nu'/\nu'_c),$
with the characteristic frequency
$\nu'_c = 3 q_e B'_{\rm rs,tot} \gamma_e'^2/(4\pi m_{\rm e} c).$
Here $F(x) = x \int_{x}^{+\infty} K_{5/3}(k)\,\mathrm{d}k$, where $K_{5/3}$ is the modified Bessel function of order $5/3$.  
By replacing the subscript “rs’’ with “fs’’ in the above expressions, one obtains the corresponding quantities for the forward-shocked region.

The IC emission includes three components: synchrotron self-Compton (SSC), external inverse-Compton (EIC) scattering of photons from the Poynting-flux jet, and EIC scattering of photons produced in the other shocked region. The comoving IC power per unit solid angle in the RS region from these three processes can be written as (\citealp{1970RvMP...42..237B,2008MNRAS.384.1483F})
\begin{equation}
	P'_{\rm rs,SSC}(\nu', R) =
	\frac{3\sigma_{\rm T} c h \nu'}{4}
	\int \frac{n'_{\rm syn,rs}(\nu'_{\rm s})}{\nu'_{\rm s}}
	\left[
	\int \frac{F(q,g)}{\gamma_{\rm e}^{\prime 2}}
	\frac{\mathrm{d}N_{\rm e,rs}}{\mathrm{d}\gamma'_{\rm e}} \, \mathrm{d}\gamma'_{\rm e}
	\right]
	\mathrm{d}\nu'_{\rm s},
\end{equation}
\begin{equation}
	P'_{\rm rs,EICjet}(\nu', R) =
	\frac{3\sigma_{\rm T} c h \nu'}{4}
	\int \frac{n'_{\rm rs,EICjet}(\nu'_{\rm s})}{\nu'_{\rm s}}
	\left[
	\int \frac{F(q,g)}{\gamma_{\rm e}^{\prime 2}}
	\frac{\mathrm{d}N_{\rm e,rs}}{\mathrm{d}\gamma'_{\rm e}} \, \mathrm{d}\gamma'_{\rm e}
	\right]
	\mathrm{d}\nu'_{\rm s},
\end{equation}
\begin{equation}
	P'_{\rm rs,EICfs}(\nu', R) =
	\frac{3\sigma_{\rm T} c h \nu'}{4}
	\int \frac{n'_{\rm syn,fs}(\nu'_{\rm s})}{\nu'_{\rm s}}
	\left[
	\int \frac{F(q,g)}{\gamma_{\rm e}^{\prime 2}}
	\frac{\mathrm{d}N_{\rm e,rs}}{\mathrm{d}\gamma'_{\rm e}} \, \mathrm{d}\gamma'_{\rm e}
	\right]
	\mathrm{d}\nu'_{\rm s},
\end{equation}
where the number densities of seed photons in the comoving frame are
$
	n'_{\rm syn,rs} (\nu'_{\rm s})
	= P'_{\rm syn,rs}(\nu'_{\rm s})/(R_{\rm rs}^2 c h\nu'_{\rm s}), 
	n'_{\rm rs,EICjet}(\nu'_{\rm s})
	= L_{\nu,{\rm jet}}(\nu'_{\rm s})/(4\pi R_{\rm cd}^2 \Gamma_{3}^2 c h\nu'_{\rm s}),
	n'_{\rm syn,fs} (\nu'_{\rm s})
	= P'_{\rm syn,fs}(\nu'_{\rm s})/(R_{\rm fs}^2 c h\nu'_{\rm s}).
$
The scattering kernel $F(q,g)$ is
\begin{equation}
	F(q,g) =
	\begin{cases}
		2q\ln q + (1 + 2q)(1 - q) + \dfrac{(qg)^2}{2(1 + qg)}(1 - q), & \text{for } 1/(4\gamma'^2_{\rm e}) < q < 1, \\[8pt]
		0, & \text{otherwise},
	\end{cases}
\end{equation}
with
$
	g = 4\gamma_{\rm e}' h\nu_{\rm s}'/(m_{\rm e} c^2),
	E = h\nu'/(\gamma_{\rm e}' m_{\rm e} c^2)$, and 
$   q = E/[g(1 - E)].$

Based on the above three IC processes, the total inverse-Compton cooling rate of electrons
in the RS region is given by
\begin{equation}\label{IC cool}
	\dot{\gamma}'_{\mathrm{e,eic}}(t_{\mathrm{obs}})
	= - \frac{1}{m_e c^2} 
	\frac{3 \sigma_{\mathrm{T}} c}{4 \gamma_e'^{\,2}}
	\int
	\frac{n'_{\rm s,tot}(\nu'_{\rm s}, t_{\mathrm{obs}})}{\nu'_{\rm s}} 
	\, d\nu'_{\rm s}
	\int_{\nu'_{\rm eic,min}}^{\nu'_{\rm eic,max}}
	F(q,g)\, h \nu'_{\mathrm{eic}} 
	\, d\nu'_{\mathrm{eic}} ,
\end{equation}
where 
$h\nu'_{\rm eic,min}=h\nu'_{\rm s}$ 
and 
$h\nu'_{\rm eic,max}
= h(\nu'_{\rm eic}/E)\,[g/(g+1)]$
are the lower and upper limits of the photon energy
for upscattered seed photons with energy $h\nu'_{\rm s}$, respectively. The total seed-photon number density is $n'_{\rm s,tot}=n'_{\rm syn,rs}+n'_{\rm rs,EICjet}+n'_{\rm syn,fs}$.

The observed total non-thermal flux from the RS region is
\begin{equation}
	F_{\rm rs,nonth}(\nu_{\rm obs}, t_{\rm obs})
	= (1+z)\, 2\pi \int_{0}^{\theta_{\rm jet}}
	\left[ f'_{\rm rs,syn}(\nu', r_{\rm obs}) + f'_{\rm rs,IC}(\nu', r_{\rm obs}) \right]
	\mathcal{D}^{3} \sin\theta \, \mathrm{d}\theta ,
\end{equation}
where
\begin{equation}
	f'_{\rm rs,syn}(\nu', r_{\rm obs}) 
	= \frac{1}{D_{\rm L}^2} 
	P'_{\rm syn,rs}(\nu', r_{\rm obs})
	\frac{1 - \exp\!\left[-(\tau_{\rm ssa,rs}+\tau)\right]}{\tau_{\rm ssa,rs}+\tau},
\end{equation}
with the synchrotron self-absorption optical depth
\begin{equation}
	\tau_{\rm ssa,rs}(\nu', r_{\rm obs})
	= - \frac{1}{8\pi m_e \nu'^2}
	\int 
	\left\{
	\gamma_e'^2 \, p'_{\rm syn,rs}(\nu', \gamma'_e)\,
	\frac{\partial}{\partial \gamma'_e}
	\left[
	\frac{1}{\gamma_e'^2} 
	\frac{\mathrm{d} N_{e,{\rm rs}}}{\mathrm{d}\gamma'_e}
	\right]
	\right\}
	\, \mathrm{d}\gamma'_e ,
\end{equation}
and
\begin{equation}
	f'_{\rm rs,IC}(\nu', r_{\rm obs})
	=  \frac{1}{D_{\rm L}^2}
	\Big[
	P'_{\rm rs,SSC}(\nu', r_{\rm obs})
	+ P'_{\rm rs,EICjet}(\nu', r_{\rm obs})
	+ P'_{\rm rs,EICfs}(\nu', r_{\rm obs})
	\Big]
	\frac{1 - e^{-\tau}}{\tau}.
\end{equation}

By replacing the subscript “rs’’ with “fs’’ in the above quantities and equations, one can straightforwardly obtain the corresponding expressions for the forward-shocked region.

\section{Results}\label{sec3}

\subsection{Dynamical evolution of the PWN}\label{sub3:1}

\begin{figure*}[htbp]
	\centering
	\begin{minipage}{0.49\linewidth}
		\includegraphics[height=0.7\textwidth, width=1\textwidth]{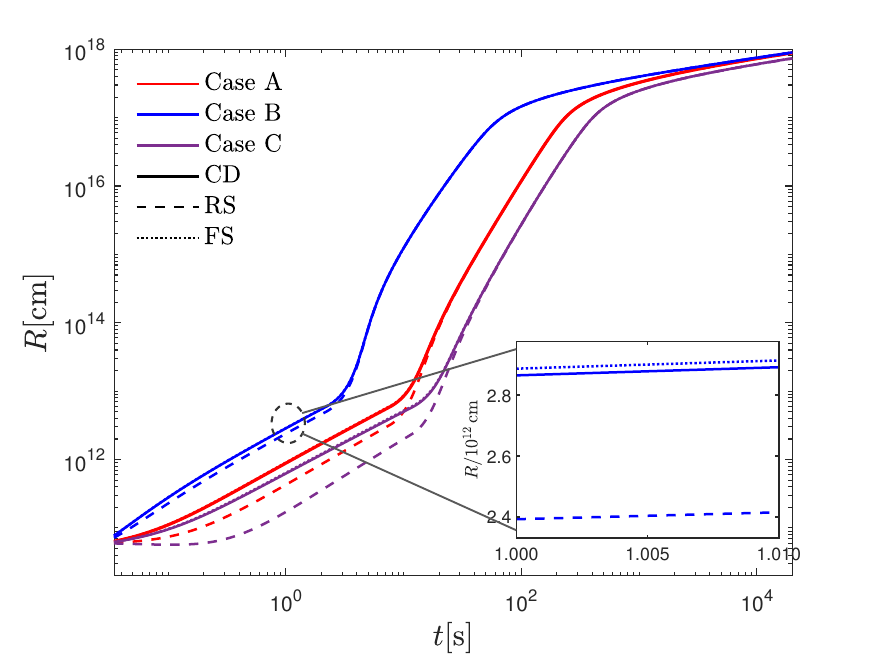}
	\end{minipage}
	\begin{minipage}{0.49\linewidth}
		\includegraphics[height=0.7\textwidth, width=1\textwidth]{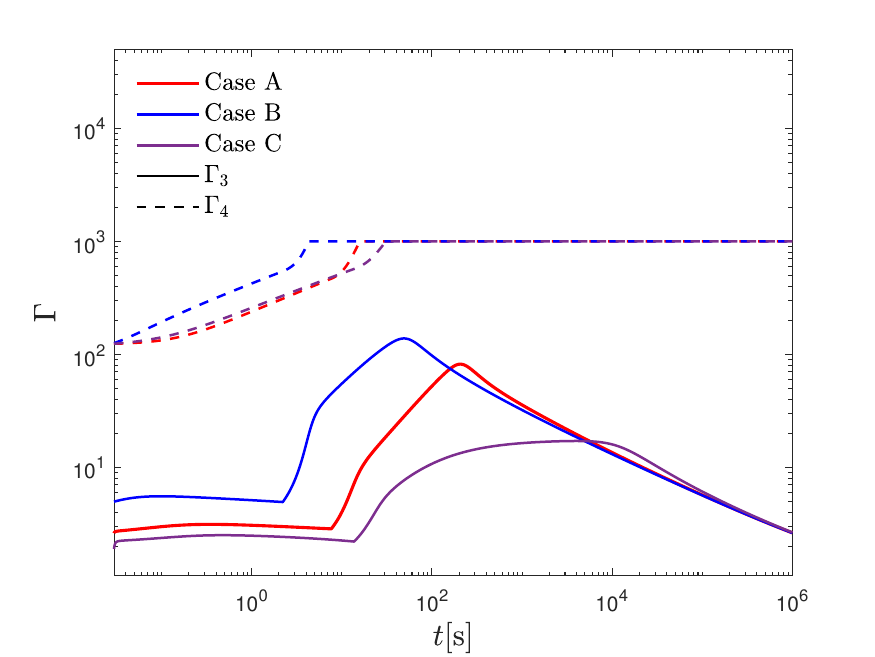}
	\end{minipage}
	\begin{minipage}{0.49\linewidth}
		\includegraphics[height=0.7\textwidth, width=1\textwidth]{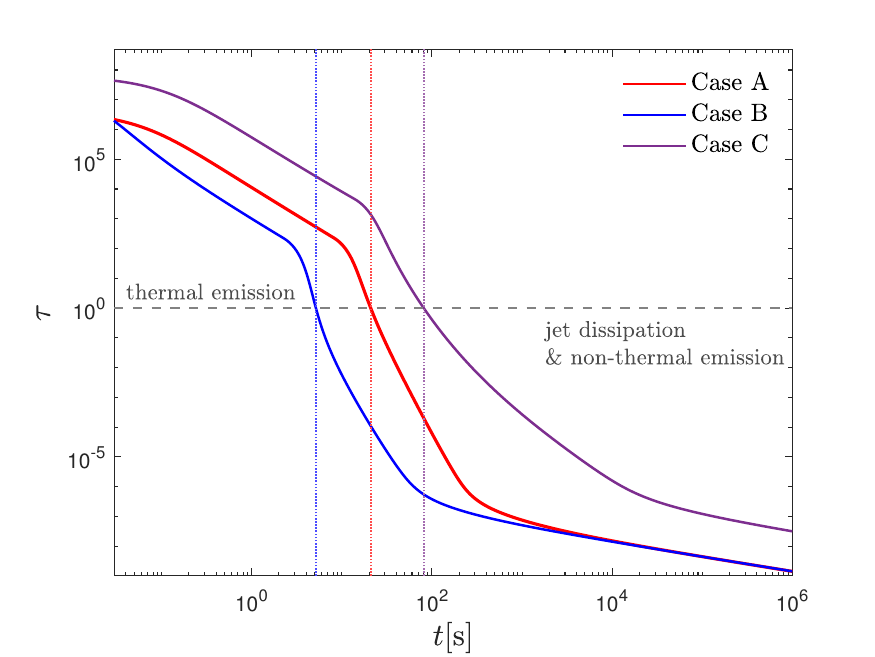}
	\end{minipage}
	\begin{minipage}{0.49\linewidth}
		\includegraphics[height=0.7\textwidth, width=1\textwidth]{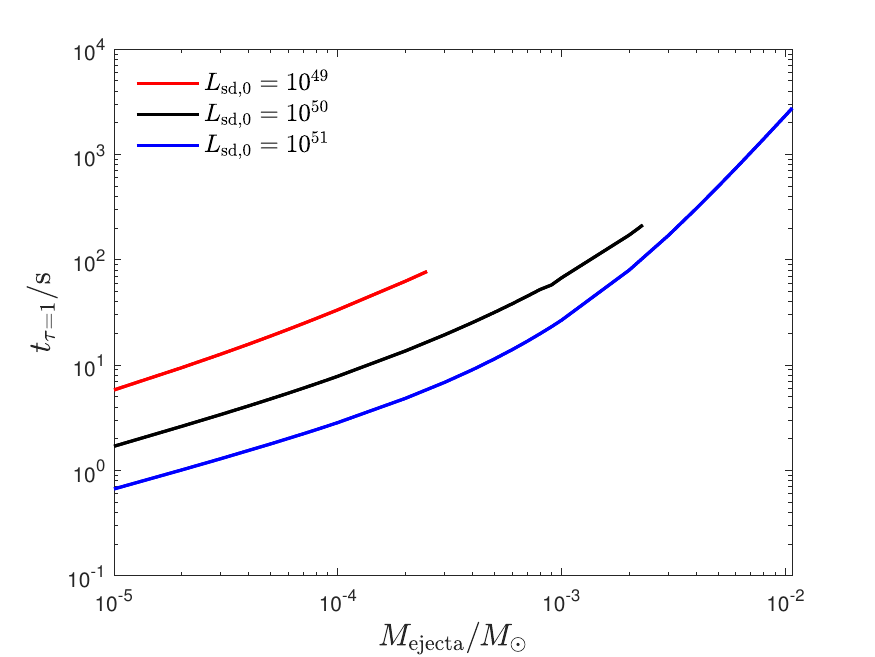}
	\end{minipage}
	\caption{
		Evolution of the dynamical quantities for different model parameters. 
		Upper left: time evolution of the radii of the contact discontinuity (solid), reverse shock (dashed), and forward shock (dotted). An inset in the upper-left panel shows a zoomed view of the radii around $t\sim 1$~s for Case~B. 
		Upper right: time evolution of the Lorentz factors in regions~3 (solid) and~4 (dashed). 
		Lower left: time evolution of the ejecta optical depth $\tau$; the horizontal dashed line marks $\tau=1$, above which the emission is dominated by thermal radiation from the ejecta, whereas below it the jet dissipation and non-thermal components dominate. The vertical dotted lines indicate the corresponding transparency times $t_{\tau=1}$ for the three cases. 
		In the upper-left, upper-right, and lower-left panels, the red, blue, and purple curves correspond to Case~A ($L_{\rm sd,0}=10^{49}\,\mathrm{erg\,s^{-1}}$, $M_{\rm ej}=10^{-4}\,M_\odot$), Case~B ($L_{\rm sd,0}=10^{50}\,\mathrm{erg\,s^{-1}}$, $M_{\rm ej}=10^{-4}\,M_\odot$), and Case~C ($L_{\rm sd,0}=10^{50}\,\mathrm{erg\,s^{-1}}$, $M_{\rm ej}=10^{-3}\,M_\odot$), respectively. 
		Lower right: transparency time $t_{\tau=1}$ as a function of ejecta mass $M_{\rm ejecta}$ for different initial spin-down luminosities, $L_{\rm sd,0}=10^{49}$, $10^{50}$, and $10^{51}\,\mathrm{erg\,s^{-1}}$.
	}
	
	\label{fig1}
\end{figure*}

We first examine the dynamical evolution of the PWN by tracking the radii of the FS, RS, and CD, together with the time evolution of the Lorentz factors in regions~3 and~4 and the optical depth of the ejecta. Figure~\ref{fig1} summarizes the results for three representative parameter sets: Case~A with $L_{\rm sd,0}=10^{49}\,\rm erg\,s^{-1}$ and $M_{\rm ej}=10^{-4}\,M_\odot$, Case~B with $L_{\rm sd,0}=10^{50}\,\rm erg\,s^{-1}$ and $M_{\rm ej}=10^{-4}\,M_\odot$, and Case~C with $L_{\rm sd,0}=10^{50}\,\rm erg\,s^{-1}$ and $M_{\rm ej}=10^{-3}\,M_\odot$.
As shown in the upper-left panel of Figure~\ref{fig1}, the RS significantly lags behind the FS and the CD until the jet passes through the ejecta. Then, the RS gradually catches up with the FS and the CD. A lower spin-down luminosity or a larger ejecta mass leads to a more pronounced lag of the RS.
An inset in the upper-left panel shows a zoomed view around $t\sim 1$~s for Case~B, illustrating the close spacing of the CD, RS, and FS radii.
The upper-right panel of Figure~\ref{fig1} shows the evolution of the Lorentz factors in regions~3 and~4. Both the spin-down luminosity and the ejecta mass significantly affect the PWN Lorentz-factor evolution: a higher $L_{\rm sd,0}$ produces an earlier and higher peak in the Lorentz factor of region~3, whereas a larger $M_{\rm ej}$ delays the peak and reduces its magnitude.

The lower-left panel of Figure~\ref{fig1} presents the temporal evolution of the ejecta optical depth $\tau$. A higher spin-down luminosity and a smaller ejecta mass cause the system to become optically thin more rapidly. The times at which $\tau=1$ for Cases~A, B, and C are approximately $20$~s, $5$~s, and $80$~s, respectively.
The lower-right panel of Figure~\ref{fig1} displays how the epoch at which $\tau=1$ depends on $M_{\rm ej}$; different colors denote different $L_{\rm sd,0}$. The cutoff in each curve marks the critical ejecta mass for inward propagation and eventual disappearance of the RS. For initial spin-down luminosities of $L_{\rm sd,0}=10^{49}$, $10^{50}$, and $10^{51}\,\rm erg\,s^{-1}$, the corresponding critical ejecta masses are $M_{\rm ej}\simeq 2.5\times10^{-4}\,M_\odot$, $2.3\times10^{-3}\,M_\odot$, and $1.07\times10^{-2}\,M_\odot$, respectively.
\citet{2022MNRAS.516.2614A} studied the dynamical evolution of engine-fed kilonovae (mergernovae) and showed that the RS in a magnetar-driven PWN can rapidly propagate inward toward the wind-acceleration region and then disappear. In our calculations, we likewise find that when $L_{\rm sd,0}$ is sufficiently low or $M_{\rm ej}$ is sufficiently large, the RS indeed propagates inward and ultimately dissipates. However, because the ejecta mass along the jet direction in neutron-star mergers is typically small ($\sim10^{-4}\,M_\odot$; \citealp{2020ApJ...901L..26R}) and post-merger magnetars are expected to possess high spin and strong magnetic fields (\citealp{2020GReGr..52..108B,2021GReGr..53...59S}), the RS is expected to survive in most cases.

\subsection{Multi-band light curve and spectrum} \label{sub3:2}

\begin{figure*}[htb]
	\centering
    \begin{minipage}{0.49\linewidth}
		\includegraphics[height=0.7\textwidth, width=1\textwidth]{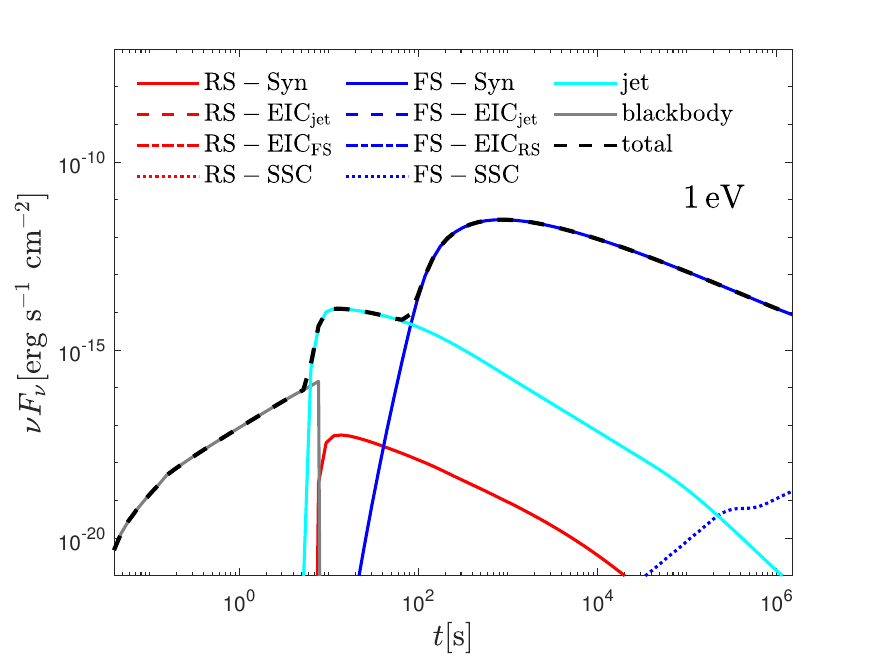}
	\end{minipage}
	\begin{minipage}{0.49\linewidth}
      \includegraphics[height=0.7\textwidth, width=1\textwidth]{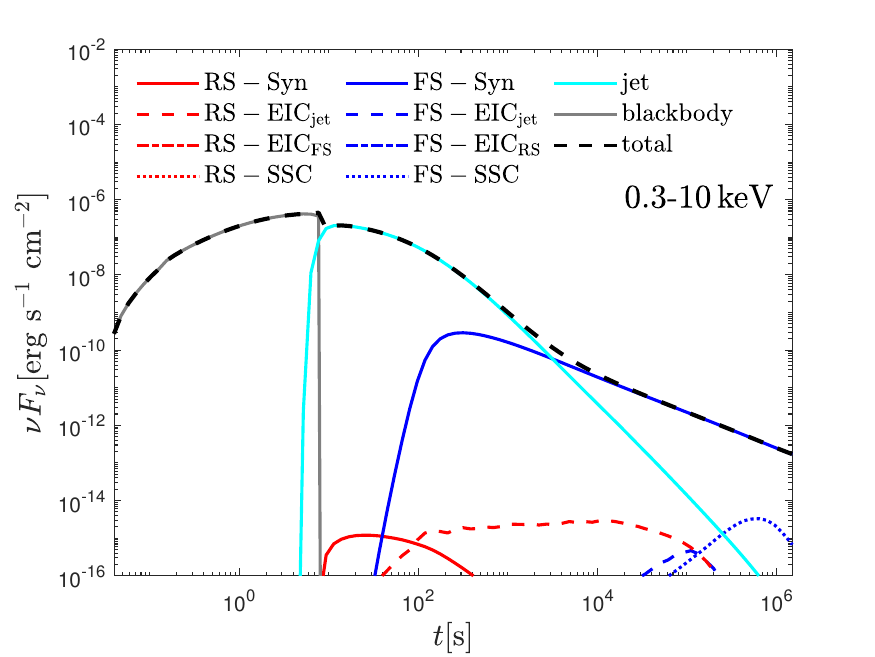}
	\end{minipage}
	\begin{minipage}{0.49\linewidth}
		\includegraphics[height=0.7\textwidth, width=1\textwidth]{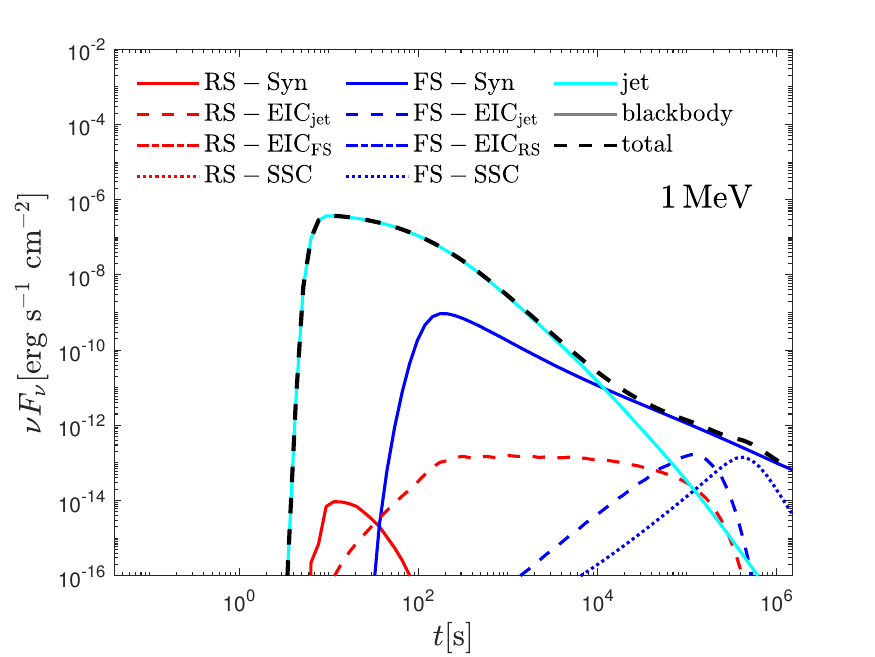}
	\end{minipage}
	\begin{minipage}{0.49\linewidth}
		\includegraphics[height=0.7\textwidth, width=1\textwidth]{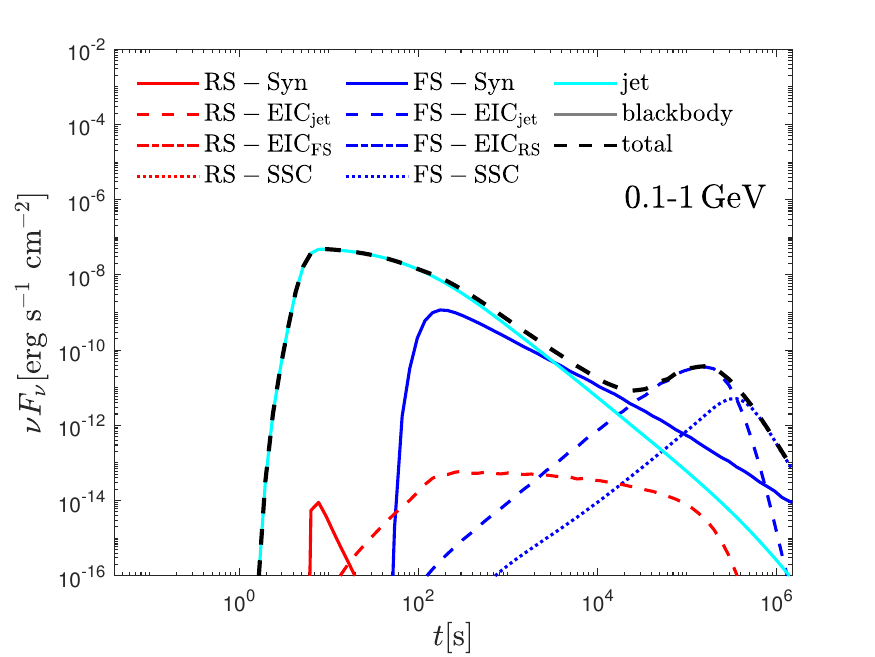}
	\end{minipage}
\caption{
	Multi-band light curves of the jet and PWN emission for Case~B. 
	From the upper left to the lower right, the panels show the light curves in the 1~eV, 0.3--10~keV, 1~MeV, and 0.1--1~GeV bands, respectively. 
	Cyan, black, red, and blue lines represent emission from the jet, ejecta blackbody, reverse shock, and forward shock, respectively. 
	Solid lines denote synchrotron radiation from the shocks. 
	Red and blue dashed lines indicate external inverse-Compton (EIC) emission in which jet photons serve as the seed photons. 
	Dash-dotted lines show EIC emission arising from cross-scattering, where seed photons from one shock region are upscattered by electrons in the other region (i.e., forward–reverse shock cross-scattering). 
	Dotted lines represent the synchrotron self-Compton (SSC) component. 
	The black dashed line shows the total emission.
}

    \label{fig2}
\end{figure*}

The neutron star formed after a binary neutron star merger is expected to be rapidly rotating and strongly magnetized, leading to a relatively high spin-down luminosity (\citealp{2020GReGr..52..108B,2021GReGr..53...59S}). 
Taking Case~B ($L_{\rm sd,0}=10^{50}\,\rm erg\,s^{-1}$, $M_{\rm ej}=10^{-4}\,M_{\odot}$) as an example, the resulting multi-band light curves from our model are shown in Figure~2. 
From the upper left to the lower right, the panels correspond to photon energies of 1~eV, 0.3--10~keV, 1~MeV, and 0.1--1~GeV, respectively.
As illustrated in Figure~\ref{fig2}, in the optical and X-ray bands, the early-time emission during the optically thick phase is dominated by the thermal blackbody radiation from the ejecta. 
Once the ejecta becomes optically thin, internal dissipation in the jet takes over as the primary emission source. 
As the jet luminosity subsequently declines, synchrotron radiation from the FS becomes dominant and exhibits a flatter decay slope. 
In the MeV and GeV bands, the blackbody component no longer contributes even during the early optically thick stage. 
After the ejecta turns optically thin, emission from internal dissipation in the jet dominates from a few seconds up to nearly $10^{4}$~s. 
In the MeV band, as the jet luminosity further decreases, synchrotron radiation from the FS becomes dominant after $\sim10^{4}$~s. 
At GeV energies, external inverse-Compton (EIC) emission from the FS, induced by scattering of the jet’s seed photons, produces a pronounced bump in the light curve at $t \gtrsim 10^{4}$~s. 
At later times, the synchrotron self-Compton (SSC) emission from the FS becomes increasingly dominant.

\begin{figure*}[htb]
	\centering
	\begin{minipage}{0.49\linewidth}
		\includegraphics[height=0.7\textwidth, width=1\textwidth]{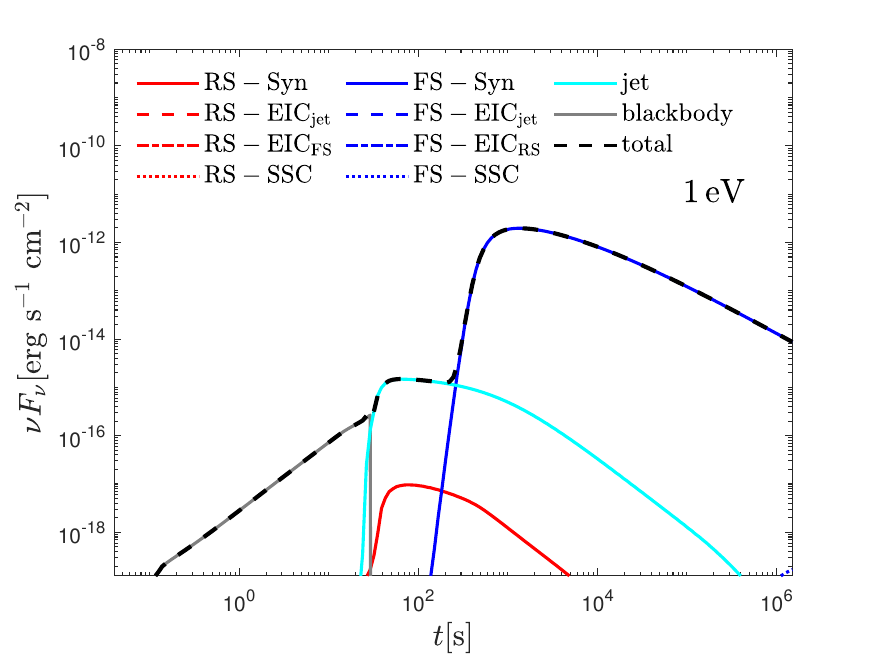}
	\end{minipage}
	\begin{minipage}{0.49\linewidth}
		\includegraphics[height=0.7\textwidth, width=1\textwidth]{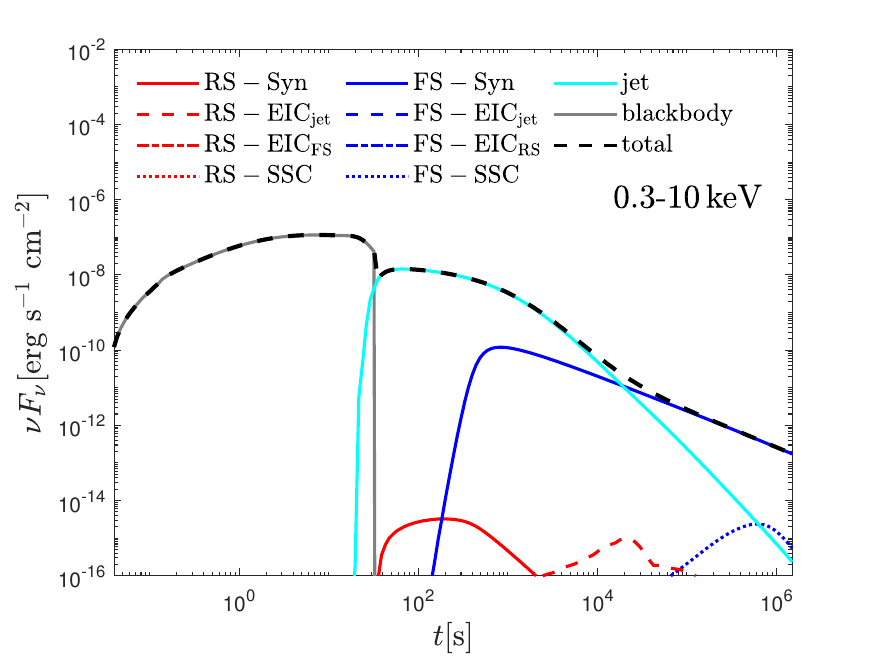}
	\end{minipage}
	\begin{minipage}{0.49\linewidth}
		\includegraphics[height=0.7\textwidth, width=1\textwidth]{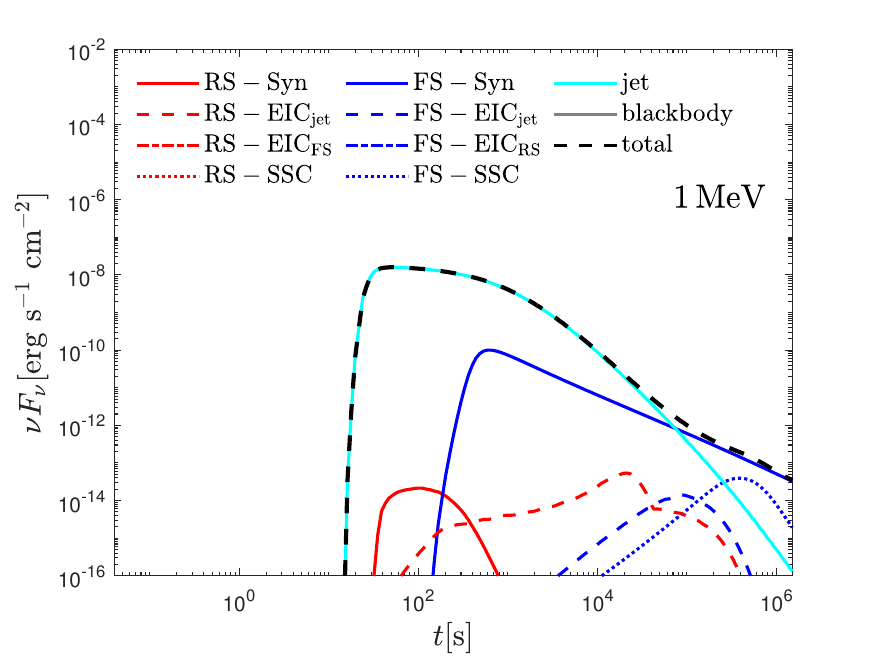}
	\end{minipage}
	\begin{minipage}{0.49\linewidth}
		\includegraphics[height=0.7\textwidth, width=1\textwidth]{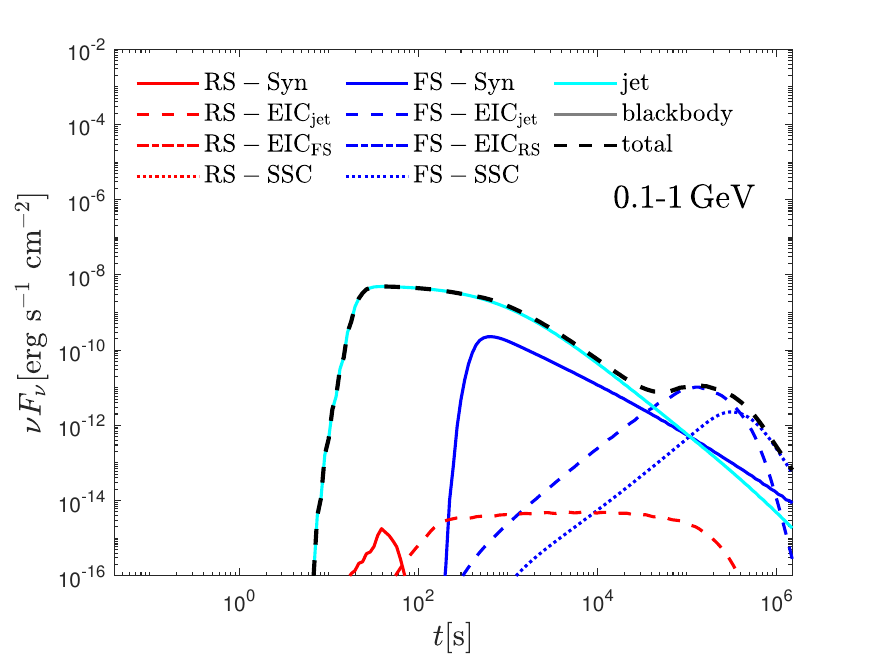}
	\end{minipage}
	\caption{Same as Figure~\ref{fig2}, but corresponding to Case~A.}
	
	\label{fig3}
\end{figure*}

For comparison, we also plot the light curve for Case~A ($L_{\rm sd,0}=10^{49}\,\rm erg\,s^{-1}$, $M_{\rm ej}=10^{-4}\,M_{\odot}$) in Figure~\ref{fig3}. 
The lower spin-down luminosity in this case leads to a longer optically thick phase, causing the ejecta’s thermal blackbody radiation to persist for a longer duration. 
In addition, the reduced jet luminosity delays the emergence of the FS synchrotron radiation and inverse-Compton (IC) components.

\begin{figure*}[htb]
	\centering
	\begin{minipage}{0.49\linewidth}
		\includegraphics[height=0.7\textwidth, width=1\textwidth]{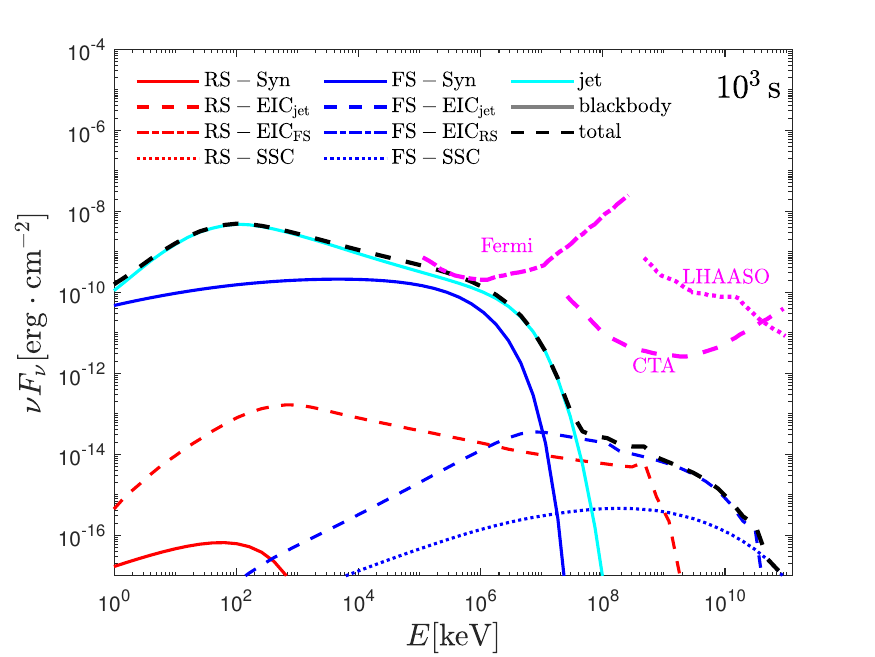}
	\end{minipage}
	\begin{minipage}{0.49\linewidth}
		\includegraphics[height=0.7\textwidth, width=1\textwidth]{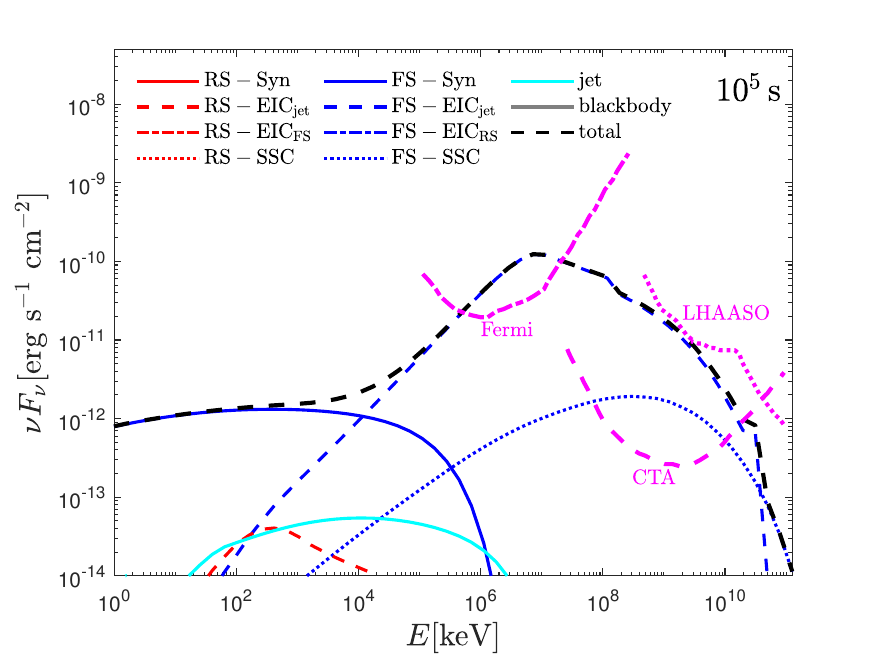}
	\end{minipage}
	
	\caption{
		Spectra of the jet and PWN emission above 1~keV for Case~B at $t = 10^3$~s (left panel) and $t = 10^{5}$~s (right panel). 
		Cyan, black, red, and blue lines correspond to emission from the jet, ejecta blackbody, reverse shock, and forward shock, respectively. 
		Solid lines denote synchrotron radiation from the shocks. 
		Red and blue dashed lines indicate external inverse-Compton (EIC) emission produced by scattering of jet photons. 
		Dash-dotted lines represent EIC emission due to cross-scattering between the forward and reverse shock regions. 
		Dotted lines show the synchrotron self-Compton (SSC) component. 
		The black dashed line represents the total emission.
        Magenta dash-dotted, dashed, and dotted lines correspond to the sensitivity curves of Fermi/LAT, CTA, and LHAASO, respectively, evaluated at $t = 10^3$~s (left) and $t = 10^5$~s (right).
	}
	
    \label{fig4}
\end{figure*}

Finally, for Case~B, we present the model spectra above 1~keV at $t = 10^3$~s and $t = 10^{5}$~s in Figure~\ref{fig4}. 
At $t = 10^3$~s, the jet emission dominates the spectrum below approximately 10~GeV. 
At higher photon energies, external inverse-Compton (EIC) emission from the forward and reverse shocks, produced by scattering of the jet’s seed photons, becomes the dominant component, although its flux is much lower than that of the jet emission at lower energies. 
By $t = 10^{5}$~s, synchrotron emission from the forward shock has replaced the jet emission as the primary component in the 1~keV–10~MeV range. 
At higher photon energies (0.01~GeV–100~TeV), EIC emission from the forward shock, with seed photons originating from the jet, becomes the dominant emission component, with a significantly higher flux than at $t = 10^3$~s.
By comparing the sensitivity curves of Fermi/LAT, LHAASO (\citealp{2019arXiv190502773C,2022ChPhC..46c0001M,2025SCPMA..6879502C}), and CTA  with the model spectra, we find that the Poynting jet–PWN system driven by a post-merger magnetar is a potentially observable source of GeV-TeV photons.

Among the observed GRB candidates of merger origin, only a few events have shown GeV-band emission, including GRB~090510, GRB~081024B, and GRB~211211A. The LAT-detected GeV emission in GRB~090510 and GRB~081024B appears on much earlier timescales ($\sim100$~s and $\sim3$~s, respectively) (\citealp{2010ApJ...716.1178A,2010A&A...524A..92C}) and is generally interpreted within the standard framework of prompt high-energy emission and/or external-shock afterglow radiation (\citealp{2010A&A...510L...7G}). To our knowledge, however, no other merger-origin GRB candidate has so far shown a late-time GeV excess on a timescale comparable to that of GRB~211211A. By contrast, GRB~211211A is the event most relevant to our model, because it exhibits longer-timescale central-engine activity in the form of long-lasting prompt/extended emission, stronger evidence for a merger origin through its associated kilonova signal, and a late-time ($\sim10^{4}$~s) GeV excess above the level expected from standard broadband afterglow models (\citealp{2022Natur.612..223R,2022Natur.612..228T,2022Natur.612..232Y,2022Natur.612..236M}).

\begin{figure*}[htb]
	\centering
	\begin{minipage}{0.49\linewidth}
		\includegraphics[height=0.7\textwidth, width=1\textwidth]{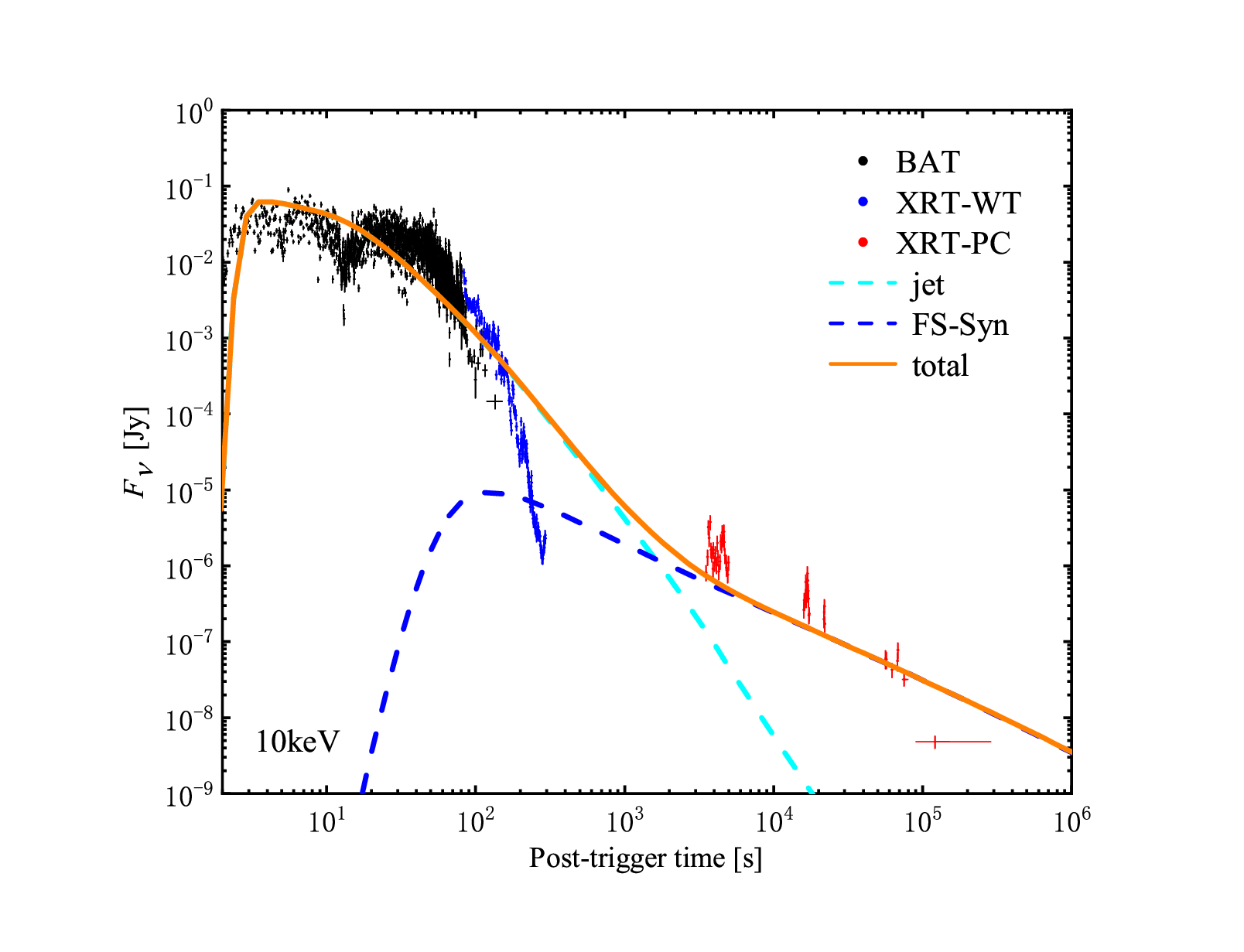}
	\end{minipage}
	\begin{minipage}{0.49\linewidth}
		\includegraphics[height=0.7\textwidth, width=1\textwidth]{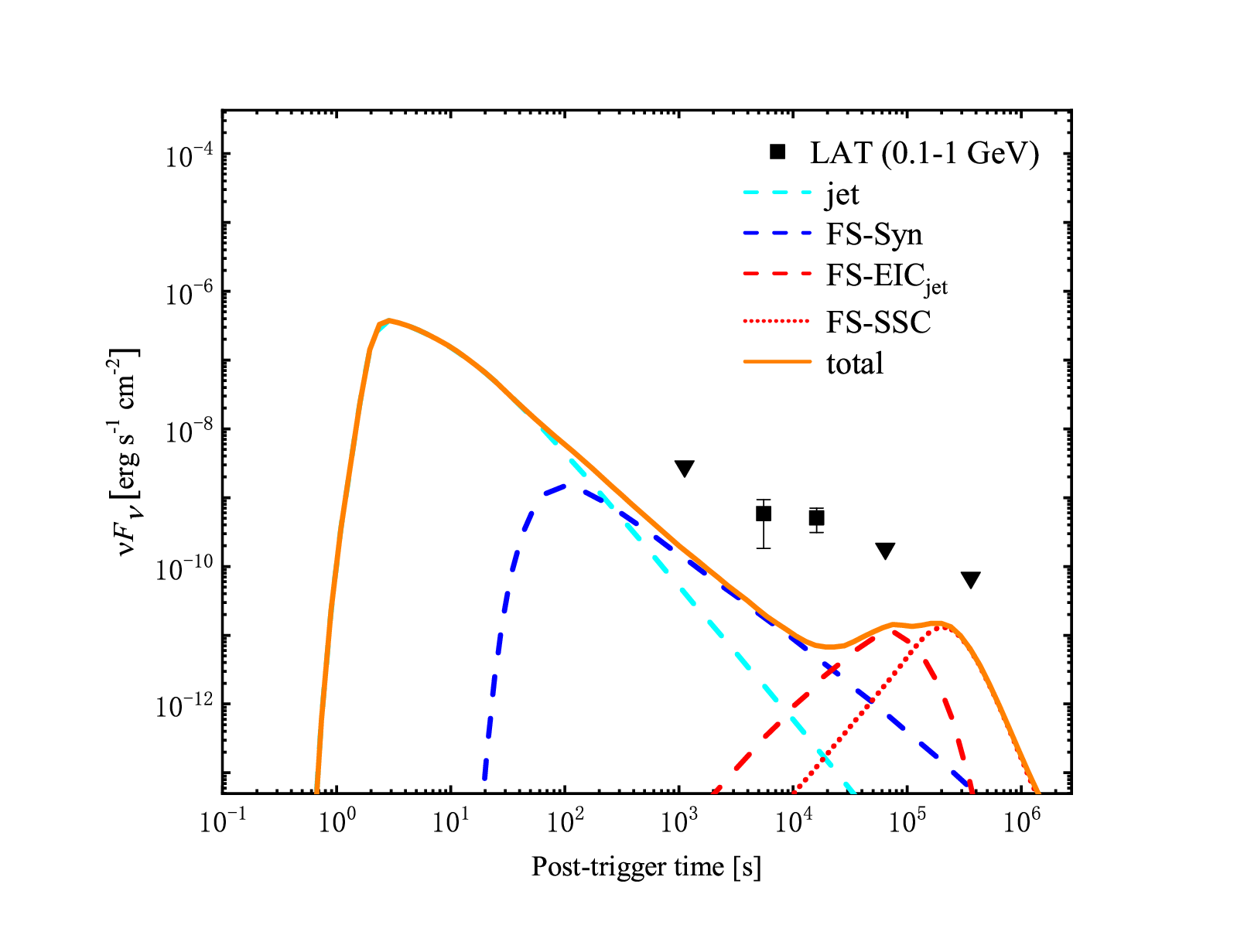}
	\end{minipage}

		\caption{
			Modeled light curves of the jet--PWN system in the 10~keV (left) and 0.1--1~GeV (right) bands for $L_{\rm sd,0}=10^{51}\,\mathrm{erg\,s^{-1}}$ and $M_{\rm ej}=10^{-4}\,M_{\odot}$, compared with the corresponding observations of GRB~211211A. 
			The cyan and blue dashed lines denote the jet emission and the forward-shock synchrotron emission, respectively, while the orange solid line shows the total model emission. 
			In the right panel, the red dashed and red dotted lines represent the forward-shock external inverse-Compton (EIC) emission produced by upscattering of jet photons and the forward-shock synchrotron self-Compton (SSC) emission, respectively. 
			The symbols show the observed data: BAT (black), XRT-WT (blue), and XRT-PC (red) in the left panel, and LAT data in the right panel; downward-pointing triangles indicate 3$\sigma$ upper limits.
		}

	\label{fig5}
\end{figure*}

We therefore compare the model light curves for an additional illustrative parameter set with $L_{\rm sd,0}=10^{51}\,\mathrm{erg\,s^{-1}}$ and $M_{\rm ej}=10^{-4}\,M_{\odot}$, which is distinct from Cases A--C, with the observed light curves of GRB~211211A in the X-ray (10~keV) and Fermi/LAT (0.1--1~GeV) bands in Figure~\ref{fig5}. We find that our model captures the broad trend of the observed X-ray light curve. In this interpretation, the jet emission is intended to represent the long-lasting prompt/extended component, whereas the forward-shock synchrotron emission corresponds to the observed afterglow. The present comparison is intended to be illustrative rather than a detailed fit. 
In particular, the steep decay at the end of the prompt/extended-emission phase is not reproduced by the current calculation and may require additional physical ingredients, such as an abrupt shutdown of the central engine, possibly caused by the collapse of a supramassive post-merger magnetar into a black hole after it loses rotational support through spin-down (\citealp{2013MNRAS.430.1061R,2014PhRvD..89d7302L,2014MNRAS.441.2433R,2022ApJ...939...51M}), together with high-latitude/curvature emission from the rapidly fading dissipation region (\citealp{2000ApJ...541L..51K,2006ApJ...642..354Z,2006ApJ...646..351L,2006MNRAS.369..311Y}).
However, the GeV bump predicted by our model, arising mainly from the forward-shock external inverse-Compton (EIC) emission with jet photons as seed photons, together with the forward-shock synchrotron self-Compton (SSC) emission, peaks later and is dimmer than the observed GeV excess of GRB~211211A. This discrepancy suggests that an additional high-energy component may be required, for example inverse-Compton scattering of kilonova photons into the GeV band (\citealp{2022Natur.612..236M}). In this picture, the EIC scattering of Poynting-jet photons by electrons accelerated in the forward shock may still contribute to a delayed GeV component, but it is unlikely to account for the entire observed GeV excess on its own.

\section{Conclusion and Discussion} \label{sec4}

The merger of binary neutron stars can produce a highly magnetized, rapidly rotating magnetar. The spin-down of such a magnetar powers a Poynting-flux–dominated jet, which interacts with the merger ejecta and the external medium and inflates a PWN. In this work, we present a self-consistent dynamical and radiative model for a magnetically dissipative Poynting jet and the associated PWN and investigate the resulting multi-wavelength emission at different evolutionary stages. 
Our calculations show that, since the high spin-down luminosities expected for post-merger magnetars and the relatively small ejecta mass along the jet direction (\citealp{2020ApJ...901L..26R,2020GReGr..52..108B,2021GReGr..53...59S}), the RS is preserved in most cases. Initially, the RS significantly lags behind both the CD and the FS. After the jet breaks out of the ejecta and enters the interstellar medium, the RS gradually catches up with the CD and FS and eventually nearly coincides with them. A lower spin-down luminosity and a larger ejecta mass both act to enhance the lag of the RS. When the spin-down power is sufficiently weak and the ejecta mass sufficiently large, the RS can even propagate inward and ultimately vanish.

The radiative behavior of the system is closely associated with the evolution of the ejecta optical depth. At early times, the ejecta are optically thick, and the optical and X-ray emission is dominated by thermal blackbody radiation from the ejecta. As the jet propagates outward and the system becomes optically thin, dissipation within the Poynting jet takes over and produces a characteristic plateau in the X-ray band. At later times, as the jet luminosity declines, synchrotron radiation from the PWN forward shock becomes the dominant contribution and exhibits a flatter decay slope. The thermal emission from the ejecta is negligible in the MeV and GeV bands. In the GeV band, the external inverse-Compton (EIC) emission from the PWN forward shock, with seed photons provided by the dissipating jet, becomes dominant and generates a pronounced bump in the light curve after $t \sim 10^{4}\,\mathrm{s}$. The model spectra further indicate that the PWN formed by the interaction between the Poynting jet and the ejecta can be an important source of TeV photons through EIC scattering at late times.
As an illustrative application, the comparison with GRB~211211A suggests that the model can capture the broad trend of its long-lasting prompt/extended emission and late-time afterglow. In particular, although the model produces a delayed GeV component, the predicted peak occurs too late and the flux is too low to account for the entire observed GeV excess, implying that an additional high-energy component, such as inverse-Compton scattering of kilonova photons, may also be required (\citealp{2022Natur.612..236M}).

Our model provides a self-consistent framework for interpreting several characteristic electromagnetic signatures of post-merger magnetar systems, including the early thermal emission, the X-ray plateau, a late-time GeV bump, and potentially observable TeV emission. Future multiwavelength observations, especially those by very-high-energy facilities such as LHAASO (\citealp{2019arXiv190502773C,2022ChPhC..46c0001M,2025SCPMA..6879502C}) and CTA (\citealp{2019scta.book.....C}), may therefore offer a powerful probe of the physical properties of magnetar remnants following BNS mergers.

\emph{Acknowledgments}
{We thank Haiming Zhang and Jia Ren for helpful discussions and the anonymous referee for constructive comments. This work is supported by the National Natural Science Foundation of China (grant Nos. 12273005, 12494575, and 12133003), the National Key R\&D Program of China (grant No. 2023YFE0117200 and 2024YFA1611700), the special funding for Guangxi Bagui Youth Scholars, and the Guangxi Talent Program (“Highland of Innovation Talents”).}

	\clearpage
	
	\clearpage
	\bibliographystyle{aasjournal}
	\bibliography{sample631}

\end{document}